\documentclass[onecolumn,preprintnumbers,amsmath,amssymb,superscriptaddress]{revtex4}
\usepackage{graphicx}
\usepackage{dcolumn}
\usepackage{bm}
\usepackage{longtable}
\usepackage{comment}
\usepackage{color}
\usepackage{float}
\usepackage[normalem]{ulem}
\newcommand{\bea}{\begin{eqnarray}}
\newcommand{\eea}{\end{eqnarray}}

\newcommand{\be}{\begin{equation}}
\newcommand{\ee}{\end{equation}}

\def\Be'{\beta_\mu^{'}}

\def\<{\bigl\langle}
\def\>{\bigr\rangle}

\usepackage{amssymb,amsfonts,amsmath,graphics,graphicx}

\begin{document}

\title{Emerging heterogeneities in Italian customs and comparison  with nearby countries}
\author{Elena Agliari}
\affiliation{Dipartimento di Matematica, Sapienza Universit\`a di Roma, Italy}
\author{Adriano Barra\footnote{Corresponding author's email: adriano.barra@roma1.infn.it}}
\affiliation{Dipartimento di Fisica, Sapienza Universit\`a di Roma, Italy}
\author{Andrea Galluzzi}
\affiliation{Dipartimento di Matematica, Sapienza Universit\`a di Roma, Italy}
\author{Marco Alberto Javarone}
\affiliation{Dipartimento di Matematica ed Informatica, Universit\`a di Cagliari, Italy}
\affiliation{Dipartimento di Scienze Umanistiche e Sociali, Universit\`a di Sassari, Italy}
\author{Andrea Pizzoferrato}
\affiliation{Mathematics Institute, University of Warwick, UK}
\author{Daniele Tantari}
\affiliation{Centro Ennio De Giorgi, Scuola Normale Superiore, Pisa, Italy}

\begin{abstract}
%
In this work we apply techniques and modus operandi typical of Statistical Mechanics to a large dataset about key social quantifiers and compare the resulting behaviours of five European nations, namely France, Germany, Italy, Spain and Switzerland. The social quantifiers considered are $i.$ the evolution of the number of autochthonous marriages (i.e. between two natives) within a given territorial district and $ii.$ the evolution of the number of mixed marriages (i.e. between a native and an immigrant) within a given territorial district.
Our investigations are twofold. From a theoretical perspective, we develop novel techniques, complementary to classical methods (e.g. historical series and logistic regression), in order to detect possible collective features underlying the empirical behaviours; from an experimental perspective, we evidence a clear outline for the evolution of the social quantifiers considered. The comparison between experimental results and theoretical predictions is excellent and allows speculating that France, Italy and Spain display a certain degree of {\em internal heterogeneity}, that is not found in Germany and Switzerland; such heterogeneity, quite mild in France and in Spain, is not negligible in Italy and highlights quantitative differences in the customs of Northern and Southern regions. These findings may suggest the persistence of two culturally distinct communities, long-term lasting heritages of different and well-established cultures.
We also find qualitative differences in the evolution of autochthonous and mixed marriages: for the former imitation in decisional mechanisms seems to play a key role (and this results in a square root relation between the number of autochthonous marriages versus the percentage of possible couples inside that country), while for the latter the emerging behaviour can be recovered (in most cases) with elementary models with no interactions, suggesting weak imitation patterns  between natives and migrants (and this translates in a linear growth for the number of mixed marriages versus the percentage of possible mixed couples in the country). However, the case of mixed marriages displays a more complex phenomenology, where further details (e.g., the provenance and the status of migrants, linguistic barriers, etc.) should also be accounted for.
\end{abstract}

\maketitle

\vskip 1cm

\section{Introduction}

In the past decade huge datasets have been captured, stored and shared to the purpose of predictive analytics. This was allowed and prompted by a number of reasons, ranging from novel techniques for massive data acquisition (especially in Biotechnological and Medical Research \cite{quake,quake2}) to the genesis of suitable repository (clouds) merging data collected from various sources/laboratories (especially in Sociological and Economical Research \cite{gabrielli,palgrave}). As a matter of fact, researchers are nowadays provided with extensive data whose hidden structures may escape classical inferential methods, hence driving the quest for proper techniques to extrapolate these inner contents.

In this context, a novel generation of Machine Learning (e.g., Deep Learning) has been devised for data mining  \cite{deep} and tools from Statistical Mechanics have been developed to extract predictive information from the available data sets.
Indeed, being firmly grounded on the law of large numbers and on the minimum energy and maximum entropy principles \cite{ellis}, Statistical Mechanics constitutes a solid approach for many applied fields, far beyond its original scope. For instance, in the last decade, it has been robustly exploited in economic complexity (see e.g., \cite{palgrave,bouchaud,stanley-PNAS,pietronero}), social complexity (see e.g., \cite{loreto,galam,barra-EPL,agliari-NJP}), and even in medicine and pharmaceutical (see e.g., \cite{barabasi,vespignani,plos-nostro}), just to cite a few.
In particular, as for the investigation of social complexity, one can rely on a bulk of stochastic techniques (see e.g.  \cite{loreto,galam,axelroad,shelling}) meant to model the underlying interacting network of the system (whose nodes are typically single decision makers and links mirror the existence of pair-wise interactions or reciprocal knowledge \cite{agliari-EPL,barra-PhysicaA}) and to figure out its emergent features. Although this perspective always implies a certain degree of simplification, its reward lies in its crucial ability to unveil {\em collective behaviors} that markedly shine over the fluctuating details.

In this paper we adopt stochastic techniques stemmed from Statistical Mechanics
to analyze extensive data regarding a key aspect of Social Complexity, that is the evolution of marriages, either between two natives (i.e., ``local marriages'') or between a native and an immigrant (i.e., ``mixed marriages''), within several European countries.

Local and mixed marriages constitute standard social quantifiers to assess, e.g., the role of
conjunctural phenomena on the family and the degree of migrant integration in a given region \cite{Bras-2008,Pennix-2008}.
The comparison between the outcomes pertaining to different countries will allow to highlight either robust (i.e. shared across the nations) or local (i.e. national) features.

In our Statistical Mechanics analysis we look at the whole population as a set of decision makers and our order parameter (namely the global observable capturing the behaviour of the system) is the density $M$ of marriages, either local or mixed. In order to estimate theoretically $M$ we refer to two prototypical models. In the former model we postulate that each decision maker acts independently of each other, being influenced only by ``external fields'' (e.g., stemming from cultural transitions or conveyed by the media). This model constitutes an adaptation of the McFadden Discrete Choice theory \cite{mcfadden} and predicts that $M$ scales linearly with the fraction $X$ of potential couples within the system, i.e., $M \propto X$.
In the other model we postulate that social interactions play a role in biasing the behaviour of decision makers and that such interactions are imitative. This model
essentially reproduces the Brock-Durlauf Discrete Choice \cite{brock} and predicts $M \propto \sqrt{X}$.
We stress that the existence of imitative social interactions within an homogeneous population is by now well evidenced and accepted; quoting Brock and Durlauf \cite{brock2,brock3}- {\em the utility an individual receives from  a given action depends directly on the choice of others in that individual's reference group}. Dealing with mixed marriages, however, this paradigma may fail as migrants can be subject to cultural differences, prejudices and discrimination, ultimately segregating them and constraining their searches within a {\em segmented marriage market} \cite{DePutte, uunk}. If this is the case, following the statistical mechanical modeling, we would expect qualitative different behaviours for the evolution of local and mixed marriages.
Indeed, as we are going to show, for local marriages the behaviour expected from interacting systems is nicely recovered, while for mixed marriages, in most cases, the behaviour expected from non-interacting systems prevails.

More precisely, in this work we will consider empirical data (available from INSEE for France, DESTATIS for Germany, ISTAT for Italy, INE for Spain, and FSO for Switzerland) and, after an extensive data analysis and model calibration, we get to the following results:\\
\begin{itemize}
\item Social interactions seem to play a major and robust role in driving local marriages since, for all the countries considered, the square-root scaling is nicely recovered; this is not the case for mixed marriages, whose evolution scales linearly with the number of available couples for most of the analyzed countries, suggesting a major role played by the independent model.

\item Focusing on local marriages, we unveil that Germany and Switzerland display internal homogeneity, that is, by repeating the analysis at a regional (rather than national) level of resolution, we find that the scaling law for local marriages is \emph{quantitatively} robust over all the regions making up each country. On the other hand, Latin countries (i.e., France, Italy, and Spain) display internal heterogeneity, namely, the scaling law for local marriages is only \emph{qualitatively} robust over all the regions making up each country. This effect is rather mild for France and Spain, but well distinguishable in Italy. In particular, in Italy one can detect two clusters of regions wherein homogeneity is recovered and, remarkably, these clusters turn out to correspond sharply to Northern and Southern regions.

\item Focusing on mixed marriages, we evidence that France and Germany essentially follow the Mc Fadden predictions, while Spain and Switzerland seem to obey the Brock-Daurlauf scenario, at least for small values of migrant's percentage inside the country and eventually loosing the imitational mechanism for higher values of migrant's percentage thus collapsing on the Mc Fadden case too. For Italy extrapolation of a clear trend is rather hard due to a very noisy behaviour, again accountable to internal heterogeneities. However, in any case, the emerging phenomenology is more complex than the one found for local marriages and we claim that further elements (e.g., the provenance and the status of migrants, linguistic barriers, etc.) play a non marginal role.
 \end{itemize}

These results are  discussed in the following sections: in Sec.~\ref{sec:results} we present the main results, corroborated by extensive data analysis; in Sec.~\ref{metodo-teorico} we show the techniques exploited, distinguishing between theoretical methodologies and data-analysis protocols; in Sec.~\ref{sec:conclusions} we summarise our results and discuss possible outlooks.

\section{Results and Discussion} \label{sec:results}
%
%
%

Our investigations aim to tackle and describe $i.$ the evolution of the density of autochthonous marriages in a given country as a function of the density of possible couples present in its population and $ii.$ the evolution of the density of mixed marriages versus the density of possible mixed couples (one native and one migrant) present in its population. The comparison between the two outcomes can shed light on similarities and differences regarding interactions between natives and among natives and migrants. Further, analysis are performed on several countries and the comparison between the related results can highlight either robust or country-dependent features.

Here, before presenting the results, we briefly describe the underlying theoretical scaffold and data-analysis methods, while we refer to the section {\em Methods} for more details and to Ref. \cite{agliari-NJP,barra-SR,agliari-EPL} for an extended treatment.

The analysis is carried out one country per time. For any given country, we first need to fix the degree of resolution at a certain territorial district, in such a way that (reasonably assuming that the number of marriages in a given district depends mainly on the population within the district itself) different districts can be considered as effectively independent realizations of the same system. We decide to fix the resolution at the provincial level, as its extent is typically broad enough to justify the hypothesis of independence \cite{agliari-NJP,barra-SR} and to wipe out local phenomena as alienation and segregation \cite{vernia}, yet the number of provinces is still large enough to get a good pool for the statistical analysis.

We call $N_P$ the total amount of provinces for the country considered ($N_P= 96$ for France, $N_P=476$ for Germany, $N_P=110$ for Italy, $N_P= 53$ for Spain, and $N_P=139$ for Switzerland, see also \cite{NUTS}), each labeled with an index $i$, namely $i=1,...,N_P$.
For each province we have time series of data concerning the fraction of males and of females, and the fraction of natives and migrants over the whole population and over a proper time window, whose extent depends on the country considered.
Data are collected yearly and we denote with $T$ the number of years sampled for the country considered ($T= 5$ [2006-2010] for France, $T=6$ [2007-2012] for Germany, $T=6$ [2005-2010] for Italy, $T= 8$ [1996,1998-2004] for Spain \footnote{In the case of mixed marriages it is $T= 6$ [1999-2004] }, and $T=3$ [2008-2010] for Switzerland), each labeled with an index $y$, namely $y=1,...,T$.

Thus, we call $\Gamma_{i, y}$ the fraction of possible couples among natives (i.e., the fraction of native males times the fraction of native females) \footnote{We did not analyze homosexual marriages as, in the countries inspected here, they were officially registered during the time window considered.} in the province $i$ and at time $y$.
Similarly, we call $\Omega_{i, y}$ the fraction of possible couples involving one native and one foreign-born (i.e., the fraction of natives times the fraction of foreign-borns) in the province $i$ and at time $y$. It is worth stressing that, as empirically evidenced in Sec.~\ref{metodo-teorico}, the ratio between males and females is constant with respect to the overall extent of the population, and this holds for both native and migrants communities in such a way that, the fraction of possible mixed, heterosexual couples is just proportional to $\Omega_{i, y}$.


The available datasets also include the time series $\{ \textrm{LM}_{i, y} \}$ and $\{ \textrm{MM}_{i,y} \}$ for, respectively, the fraction of local marriages and of mixed marriages with respect to the overall number of marriages occurred in the province $i$ at time $y$; again, we refer to Sec.~\ref{metodo-teorico} for a detailed definition.

In the following treatment, for simplicity, we will generically denote with $X_{i, y}$ and with $M_{i, y}$ the number of possible couples and of effective marriages in the province $i$ in the year $y$; according to the phenomenon we mean to model, $X_{iy}$ and $M_{iy}$ will be replaced by $\Gamma_{iy}$ and $\textrm{LM}_{iy}$ (dealing with local marriages) or by $\Omega_{iy}$ and $\textrm{MM}_{iy}$ (dealing with mixed marriages).


Each agent making up the system is looked at as a decision maker, the decision being whether to contract or not contract a marriage with another agent. There are two prototypical models for such a system of decision makers, one is close in spirit to the McFadden Discrete Choice Theory \cite{mcfadden}, the other is close to the Brock-Daurlauf theory \cite{brock,brock3}.
More precisely, in the former model one assumes that each agent decides independently of the other agents (i.e. one-body model), while in the latter one assumes that social interactions are present and each agent is influenced by the choice of the other agents (i.e. an imitative two-body model). These two opposite scenarios constitute the {\em extremal cases}, such that the related predictions provide bounds for the evolution of the average number of marriages within a country. The predictions for the two models are briefly recalled hereafter (see Sec.~\ref{metodo-teorico} for more details):
\begin{itemize}
\item Model with no interactions [McFadden scenario]\\
The amount of marriages $M_{i, y}$, within the province $i$,  scales linearly with the fraction of possible couples $X_{i, y}$, namely $M_{i, y} \propto  X_{i, y}$ \cite{agliari-NJP,barra-SR}.
Otherwise stated, as the time goes by, $X_{i,y}$ changes and $M_{i,y}$ changes too, but in a correlated way, that is according to a linear law.
Under the assumption of homogeneity, the same holds for the overall number of marriages $M_y$ at the country level
\be\label{scalingM}
M_y \sim X_y,
\ee
where $X_y$ is the fraction of possible couples in the whole country at time $y$, while $M_y$ is the fraction of effective marriages celebrated at time $y$ over the whole country.

\item Model with social interactions [Brock and Durlauf scenario]\\
The amount of marriages $M_{i, y}$ within the province $i$ scales with the square-root of
$X_{i,y}$, namely $M_{i, y} \propto  \sqrt{X_{i, y}}$ \cite{agliari-NJP,barra-SR}. Otherwise stated, as the time goes by, $X_{iy}$ changes and $M_{iy}$ changes in a correlated way, this time according to a square-root law.
Under the assumption of homogeneity, the same holds for marriages $M_y$ at the country level
\be\label{scaling}
M_y \sim \sqrt{X_y}.
\ee
\end{itemize}


We now proceed with the data analysis and the comparison with the previous theoretical predictions (i.e., Eqs.~\ref{scalingM} and \ref{scaling}), treating separately the case of local and mixed marriages.

\begin{figure}[htb!]
\begin{center}
\includegraphics[scale=0.4]{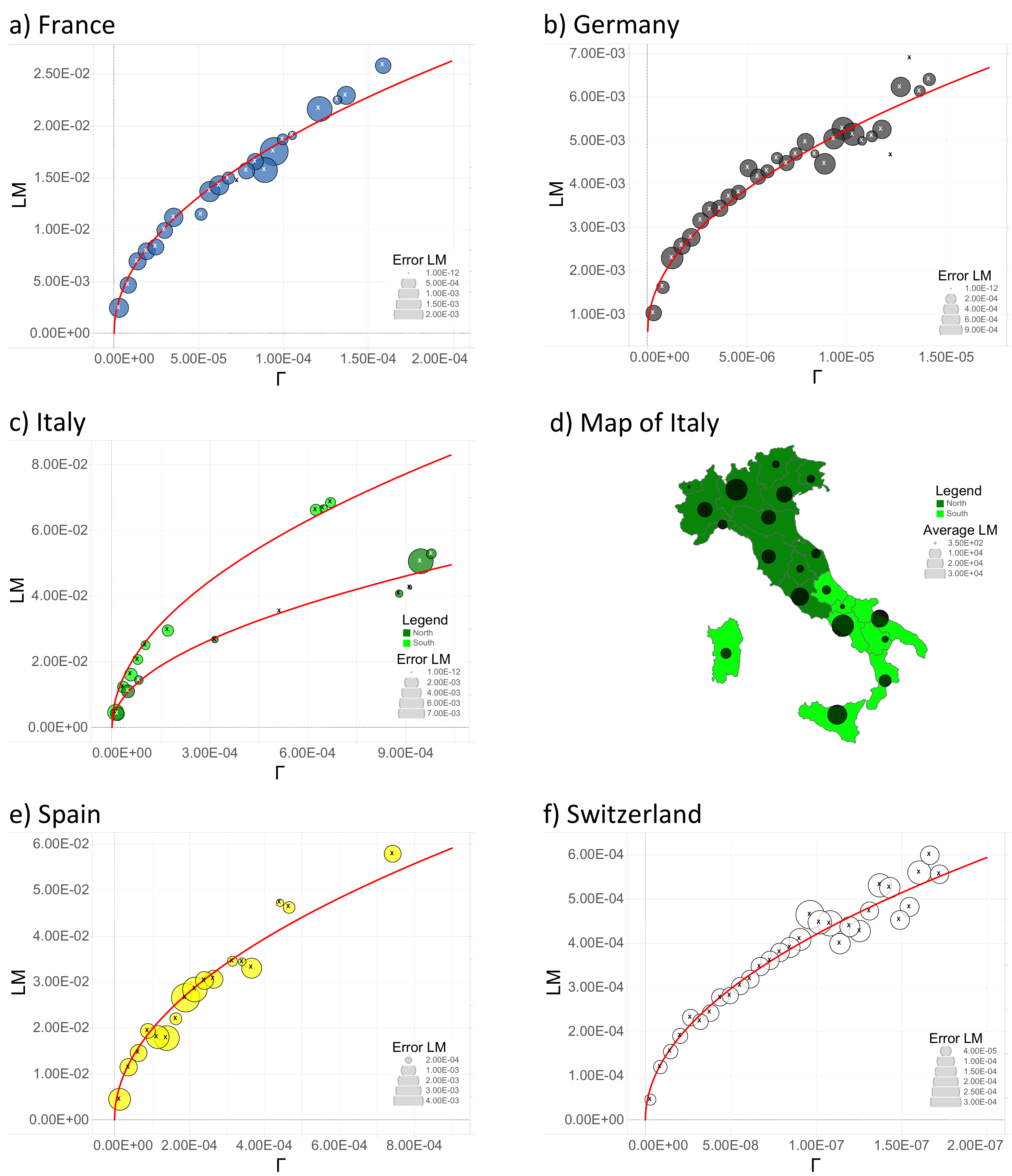}
\caption{Local marriages LM versus the fraction of potential available couples $\Gamma$ for France (panel $a$), Germany (panel $b$), Italy (panel $c$ and $d$), Spain (panel $e$), and Switzerland (panel $f$). The data points (circles) presented here are obtained from the raw data $\textrm{LM}_{i,y}$ and $\Gamma_{i,y}$ representing, respectively, the number of local marriages in the province $i$ at time $y$ divided by the total number of marriages. Details on the manipulation of raw data are provided in Sec.~\ref{metodo-teorico}.
The error circles should be understood with respect to the legend reported in each graph. In order to establish a direct comparison among the states, all the plots  were realized dividing the interval of data in $30$ bins. For each data set (circles) the best fit(s) according to Eq.~\ref{scaling} is also provided. As it is clear from the plots, and numerically confirmed by the tables presented in Fig.~\ref{fig:table}, Italy is best fitted by two square root curves that, remarkably, naturally split the country exactly into Northern and Southern regions as reported by Eurostat.
The two-color map shown for Italy depicts the definitions of Northern and Southern Italy as given by the European Union (Eurostat data) and that perfectly matches results from our data analysis. The black circles, whose sizes mirror the related amount of available data, represent examples of cities whose marriages along the years 2000-2010 have been studied.}
\label{fig:local}
\end{center}
\end{figure}
\begin{figure}[htb!]
\begin{center}
\includegraphics[scale=0.25]{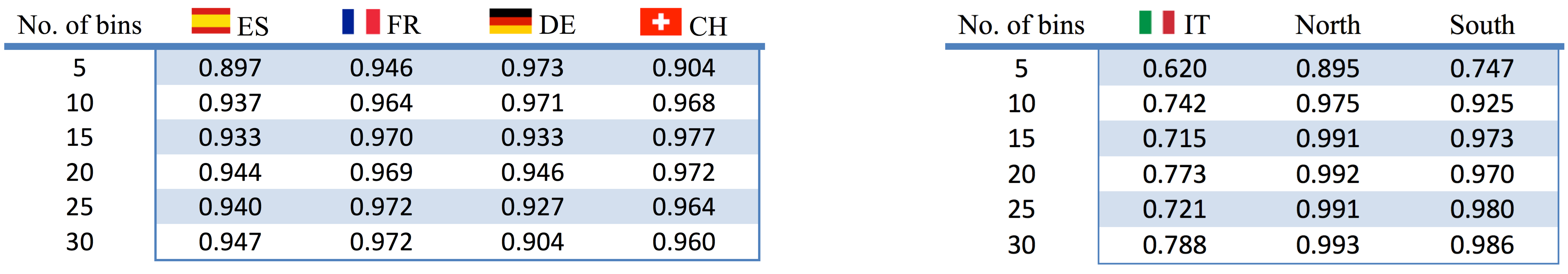}
\caption{These two tables show values of $R^2$ coupled to the best fits of data sets $\textrm{LM}_{i,y}$  versus $\Gamma_{i,y}$ for all the countries considered and for several choices of binning; the case with $30$ bins is the one shown in Fig.~\ref{fig:local}. As for Italy, whatever the level of resolution we fix (i.e. whatever the amount of bins we use), the best fit considering the country as a whole is always significantly worst than the one obtained considering Northern and Southern regions separately.}
\label{fig:table}
\end{center}
\end{figure}

\subsection{Analysis of local marriages}

In this section we focus on the data analysis of local marriages (i.e., marriages between two natives) within France, Germany, Italy, Spain and Switzerland starting from the historical series collected.
\newline
As shown in Figure \ref{fig:local}, the behavior predicted by the model with social interactions (eq. \ref{scaling}) successfully  matches data for all the countries considered but Italy. In fact, if analyzed as a single, homogeneous set, data for Italy are too noisy to detect any clear trend (see also Figure \ref{fig:table}), yet, by treating data for Northern Italy and for Southern Italy separately (as if they were two different countries), the expected square-root behavior clearly emerges.
\newline
It is worth noticing that, to obtain this result, we split Italy in such a way that the goodness of the fit (measured in terms of relative $R^2$) is maximal, and this division turns out to coincide with the definition of Northern and Southern Italy as reported by Eurostat.

In order to deepen this point and check whether ``hidden heterogeneities'' also occur in the other countries, we perform further analysis, where an intermediate level of resolution is introduced. More precisely, each province is now associated to a region, which constitutes a coarser territorial district. We call $N_R$ the total amount of regions for the country considered ($N_R = 22$ for France, $N_R = 16$ for Germany, $N_R = 20$ for Italy, $N_R = 18$ for Spain, $N_R = 26$ for Switzerland), each labeled with an index $\alpha$, and repeat the analysis region by region. As a results, the data series are reshuffled as $X_{\alpha_i, y}$ and $M_{\alpha_i, y}$, where $\alpha_i$ denotes the province $i$ in the region $\alpha$; in general, each region $\alpha$ include a different number $N_{P_{\alpha}}$ of provinces, in such a way that $\alpha_i = 1, ..., N_{P_{\alpha}}$.

\begin{figure}[htb!]
\begin{center}
\includegraphics[scale=0.63]{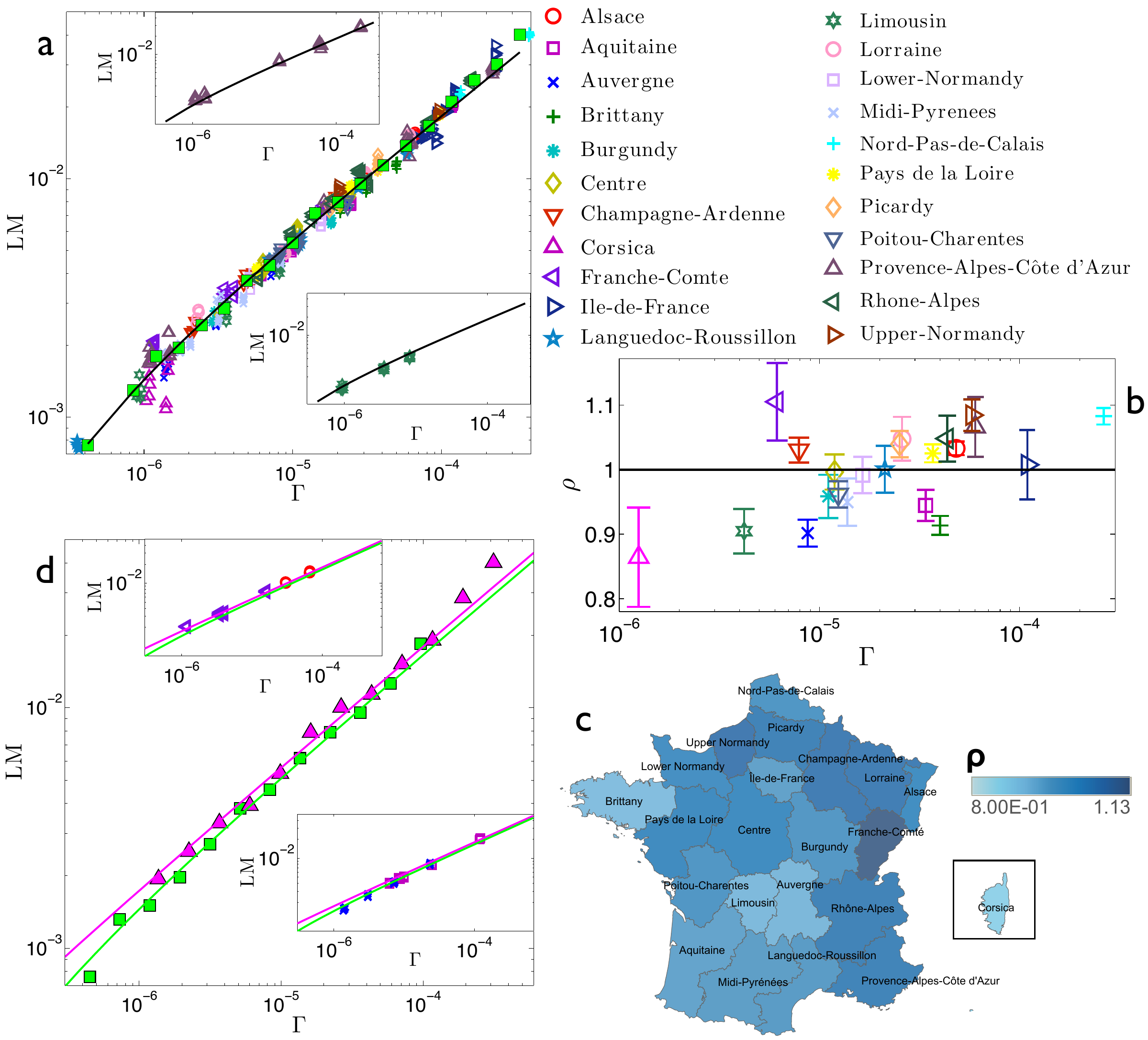}
\caption{Refined data analysis for local marriages in France. Panel $a$: log-log scale plot of $\textrm{LM}_{\alpha_i,y}$ versus $\Gamma_{\alpha_i,y}$, where different regions are denoted in different colours and symbols, as explained in the legend. These data are properly binned (green squares) and best-fitted (solid line) by $y(\Gamma) = a \sqrt{\Gamma} + b$, in agreement with the theoretical result (\ref{scaling}). The best-fit coefficients are $a=1.87$ and $b=-4.38 \cdot 10^{-4}$ with $R^2=0.97$. The parameter $b$ is introduced to account for the error (calculated in terms of the standard deviation) associated to binned data, which is $\approx 5 \%$.
In general, the various regions seem to be homogeneously scattered around the best-fit curve.
In the insets we show, as examples, the data pertaining to two particular regions, namely Limousin (upper inset) and Provence-Alpes-Cote d'Azur (lower inset). Notice that for both the insets, the best-fit previously obtained for the whole data set (solid line) still provides a proper fit.
Panel $b$: For each region we calculate $\rho_{\alpha}$, as defined in the text and deepened in Sec.~\ref{metodo-teorico}. The horizontal line is drawn as a reference for the unitary value. Notice that the largest deviation from the unitary value is for Corsica. From this plot we can distinguish regions displaying a relatively large number of marriages (i.e., $\rho_{\alpha}>1$) and regions displaying a relatively small number of marriages (i.e., $\rho_{\alpha}<1$). This division is highlighted in the colormap presented in panel $c$. Interestingly, regions exhibiting analogous deviations share a certain degree of geographical proximity: regions with $\rho_{\alpha}>1$ correspond to the North-Eastern border of France, while regions with $\rho_{\alpha}<1$ corresponds to the Center-Western part of France.
Panel $d$: the two clusters of regions highlighted are analysed separately. For each we binned the related raw data and get a best fit, still according to the function $y(\Gamma) = a \sqrt{\Gamma} + b$, obtaining $a_{up} = 1.90$ and $b_{up} =-6.5 \cdot 10^{-5}$ ($R^2=0.97$) for the set of regions with $\rho_{\alpha}>1$, and $a_{down} = 1.68$ and $b_{down} =-2.26 10^{-4}$ ($R^2=0.98$) for the set of regions with $\rho_{\alpha}<1$; notice that $a_{up}/a_{down} \approx 1.1$. Binned data for the former set (triangles) and for the latter set (square) are shown in the main panel, together with the related best fits, in a log-log scale plot.
These fits are slightly better that the one obtained at the country level, suggesting that possible internal heterogeneities may be  rather limited. In the insets we compare these best fits with raw data for two regions (Aquitaine and Auvergne) with $\rho_{\alpha}>1$ (upper inset) and two regions (Alsace and Franche-Comt\'e) with $\rho_{\alpha}< 1$ (lower inset). Notice that in both cases data points overlap both curves, again suggesting that the division highlighted here is rather mild.}
\label{fig:FRnat}
\end{center}
\end{figure}

\begin{figure}[htb!]
\begin{center}
\includegraphics[scale=0.6]{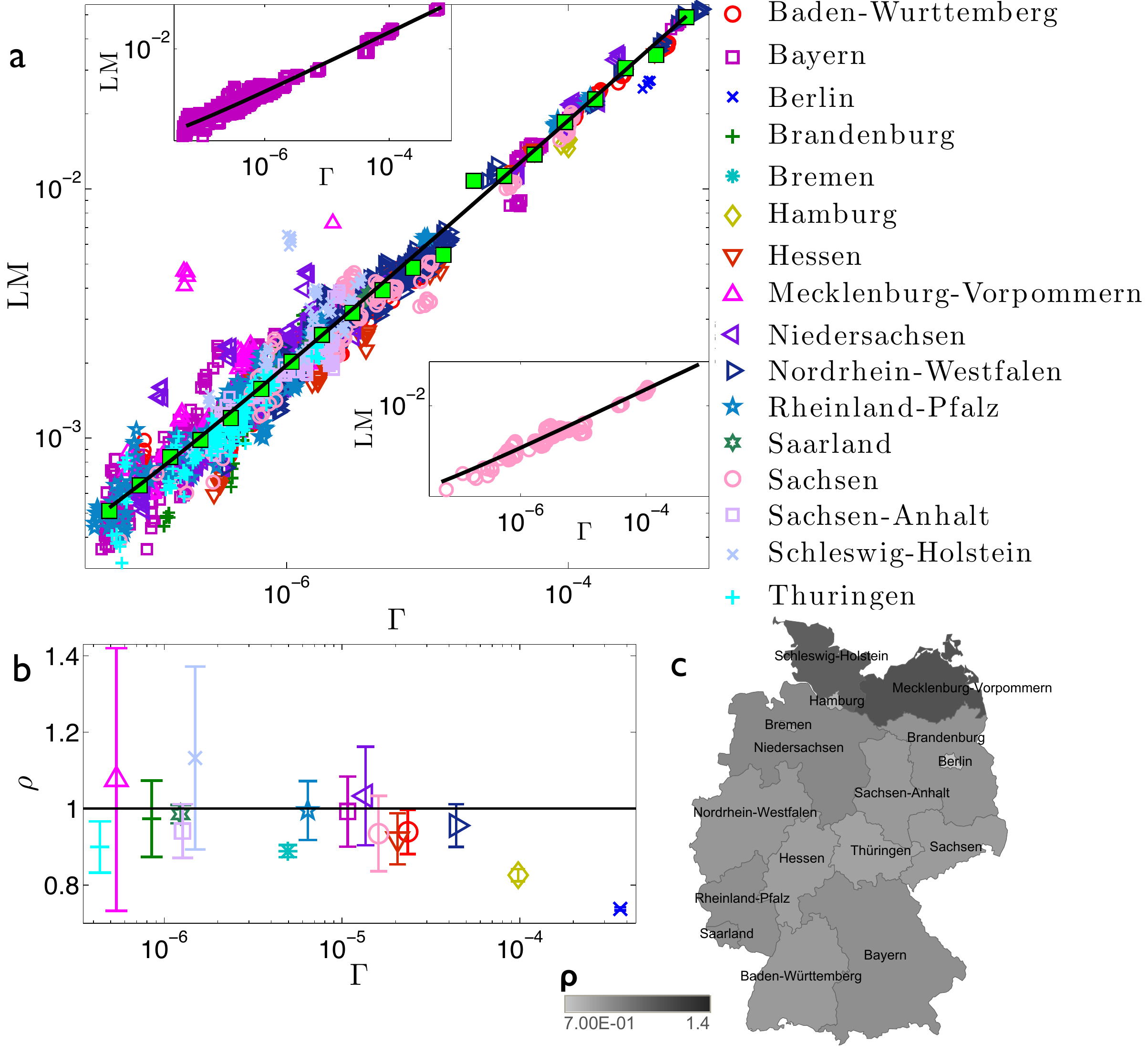}
\caption{Refined data analysis for local marriages in Germany. Panel $a$: log-log scale plot of $\textrm{LM}_{\alpha_i,y}$ versus $\Gamma_{\alpha_i,y}$, where different regions are denoted in different colours and symbols, as explained in the legend.
These data are properly binned (green squares) and best-fitted (solid line) by $y(\Gamma) = a \sqrt{\Gamma} + b$, in agreement with the theoretical result (\ref{scaling}). The best-fit coefficients are $a=1.87$ and $b=8.62 \cdot 10^{-5}$ with $R^2=0.99$.
Notice that the parameter $b$ accounts for the standard deviation associated to binned data. 
In general, the various regions seem to be homogeneously scattered around the best-fit curve.
In the insets we show, as examples, the data pertaining to two particular regions, namely Bayern (upper inset) and Sachsen (lower inset). Notice that for both the insets the best-fit previously obtained for the whole data set (solid line) still provides a very good fit.
Panel $b$: For each region we calculate $\rho_{\alpha}$, as defined in the text and deepened in Sec.~\ref{metodo-teorico}. The horizontal line is drawn as a reference for the unitary value. Notice that for most regions the deviation with respect the unitary value is within the error bar. The largest deviation from the unitary value is for two city-states, namely Hamburg and Berlin. Given that all the regions (a few cases apart) are compatible with the overall best fit no further analysis on territorial homogeneity is performed. The colormap presented in panel $c$ highlights the deviation of $\rho_{\alpha}$ with respect to the unitary value.}
\label{fig:DEnat}
\end{center}
\end{figure}

\begin{figure}[htb!]
\begin{center}
\includegraphics[scale=0.6]{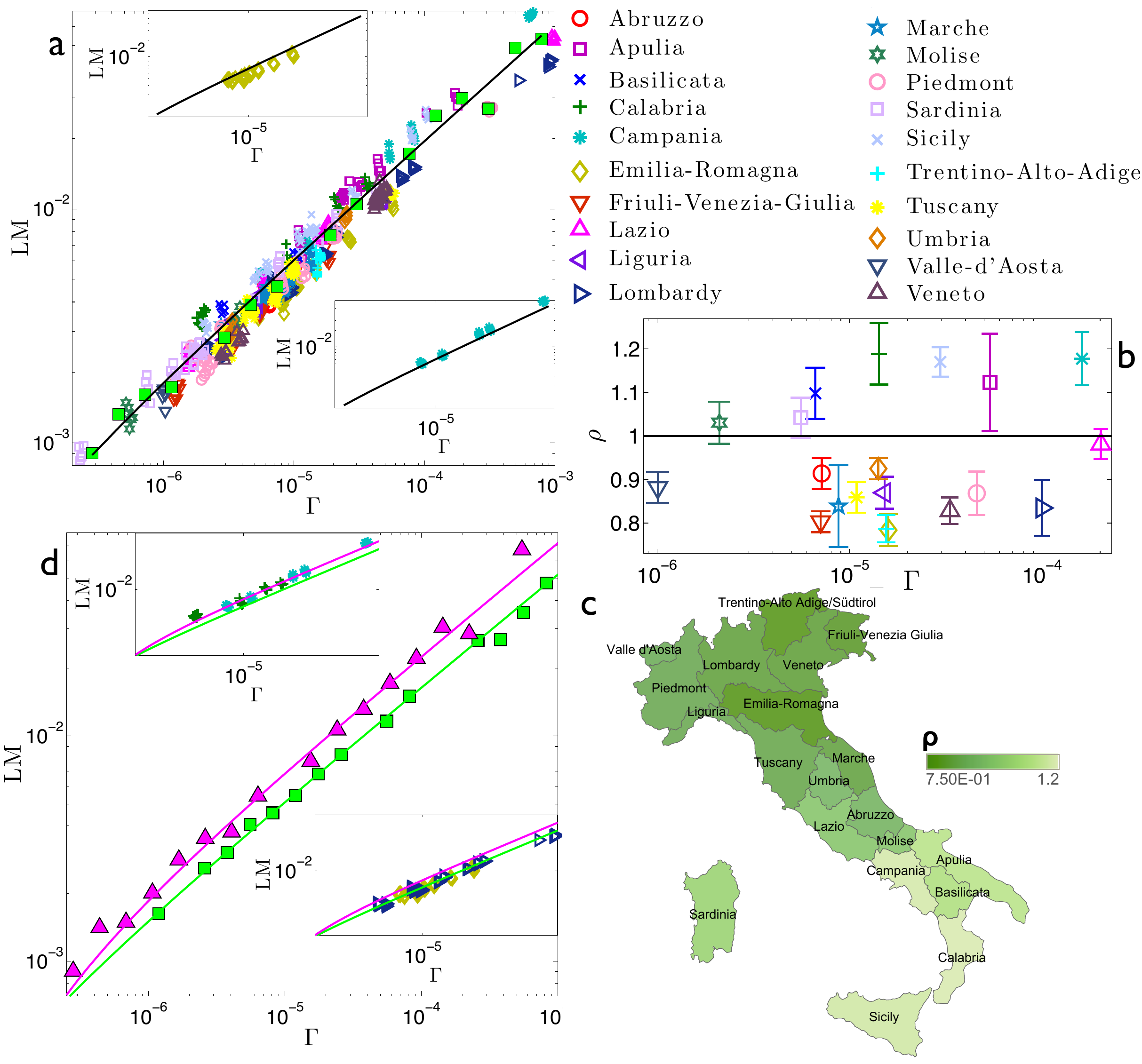}
\caption{Refined data analysis for local marriages in Italy. Panel $a$: log-log scale plot of $\textrm{LM}_{\alpha_i,y}$ versus $\Gamma_{\alpha_i,y}$, where different regions are denoted in different colours and symbols, as explained in the legend.
As discussed in the text, these data are particularly noisy, yet one can bin the whole set of data (green squares) and get the best-fitted (solid line). This is given by the function $y(\Gamma) = a \sqrt{\Gamma} + b$, in agreement with the theoretical result (\ref{scaling}). The best-fit coefficients are $a=1.96$ and $b=-1.59 \cdot 10^{-4}$ with $R^2=0.97$.
Notice that the parameter $b$ accounts for the standard deviation associated to binned data. 
It is clear, even by visual inspection, that most regions systematically deviate from the best-fit curve.
In the insets we show, as examples, the data pertaining to two particular regions, namely Emilia-Romagna (upper inset) and Campania (lower inset). Notice that the related data lay, respectively, below and above the best-fit curve previously obtained for the whole data set (solid line).
Panel $b$: For each region we calculate $\rho_{\alpha}$, as defined in the text and deepened in Sec.~\ref{metodo-teorico}. The horizontal line is drawn as a reference for the unitary value. Notice that for most regions the deviation with respect the unitary value is significant. From this plot we can distinguish regions displaying a relatively large number of marriages (i.e., $\rho_{\alpha}>1$) and regions displaying a relatively small number of marriages (i.e., $\rho_{\alpha}<1$). This division is highlighted in the colormap presented in panel $c$. Remarkably, regions exhibiting analogous deviations share a sharp geographical proximity: regions with $\rho_{\alpha}>1$ corresponds to Southern Italy, while regions with $\rho_{\alpha}<1$ corresponds to Northern Italy.
Panel $d$: the two clusters of regions highlighted are analysed separately. For each we binned the related raw data and get a best fit, still according to the function $y(\Gamma) = a \sqrt{\Gamma} + b$, obtaining $a_{up} = 2.29$ and $b_{up} =-4.39 \cdot 10^{-4}$ ($R^2=0.98$) for the set of regions with $\rho_{\alpha}>1$, and $a_{down} =1.66$ and $b_{down} =-1.52 \cdot 10^{-4}$ ($R^2=0.99$) for the set of regions with $\rho_{\alpha}<1$; notice that $a_{up}/a_{down} \approx 1.4$. Binned data for the former set (triangles) and for the latter set (square) are shown in the main panel, together with the related best fits, in a log-log scale plot.
These fits are significantly better that the one obtained at the country level, suggesting the existence of internal heterogeneities. In the insets we compare these best fits with raw data for two regions (Calabria and Campania) with $\rho_{\alpha}>1$ (upper inset) and two regions (Emilia-Romagna and Lombardy) with $\rho_{\alpha}< 1$ (lower inset). Notice that the two sets of data overlap only the pertaining fitting curve, further suggesting that the division highlighted here is not negligible.}
\label{fig:ITnat}
\end{center}
\end{figure}

\begin{figure}[htb!]
\begin{center}
\includegraphics[scale=0.61]{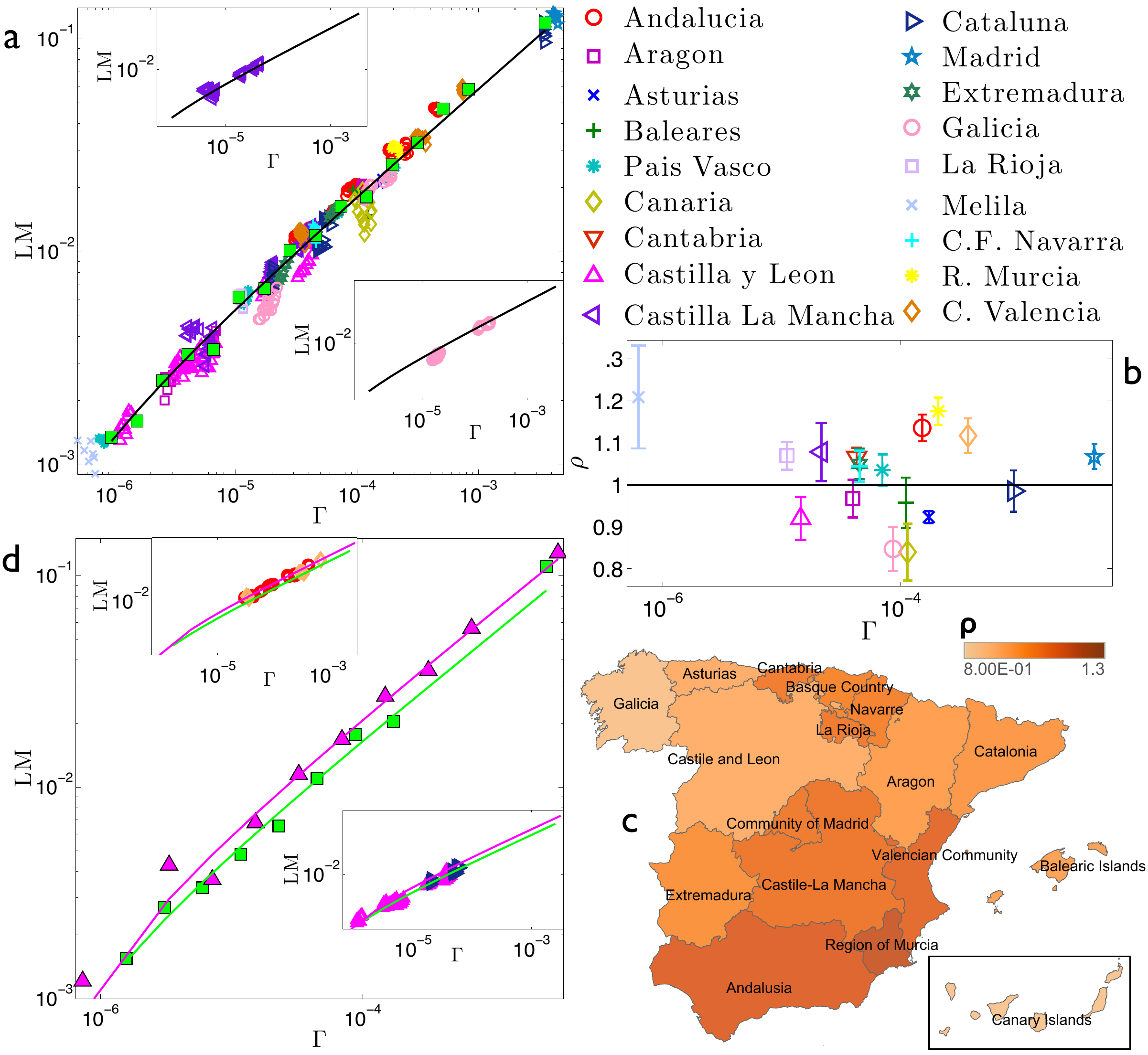}
\caption{Refined data analysis for local marriages in Spain. Panel $a$: log-log scale plot of $\textrm{LM}_{\alpha_i,y}$ versus $\Gamma_{\alpha_i,y}$, where different regions are denoted in different colours and symbols, as explained in the legend. These data are properly binned (green squares) and best-fitted (solid line) by $y(\Gamma) = a \sqrt{\Gamma} + b$, in agreement with the theoretical result (\ref{scaling}). The best-fit coefficients are $a=1.852$ and $b=-5.16 \cdot 10^{-4}$ with $R^2=0.99$.
Notice that the parameter $b$ accounts for the standard deviation associated to binned data. 
It is clear, even by visual inspection, that a few regions slightly deviate from the best-fit curve.
In the insets we show, as examples, the data pertaining to two particular regions, namely Castilla La Mancha (upper inset) and Galicia (lower inset). Notice that for both the best-fit previously obtained for the whole data set (solid line) still provides a proper fit. 
Panel $b$: For each region we calculate $\rho_{\alpha}$, as defined in the text and deepened in Sec.~\ref{metodo-teorico}. The horizontal line is drawn as a reference for the unitary value. Notice that the largest deviation from the unitary value is for Melila. From this plot we can distinguish regions displaying a relatively large number of marriages (i.e., $\rho_{\alpha}>1$) and regions displaying a relatively small number of marriages (i.e., $\rho_{\alpha}<1$). This division is highlighted in the colormap presented in panel $c$. Interestingly, regions exhibiting analogous deviations share a certain degree of geographical proximity: regions with $\rho_{\alpha}>1$ corresponds to the Southern part of Spain and to Basque region (with some adjacent regions), while regions with $\rho_{\alpha}<1$ corresponds to the North-Western part of Spain.
Panel $d$: the two clusters of regions highlighted are analysed separately. For each we binned the related raw data and get a best fit, still according to the function $y(\Gamma) = a \sqrt{\Gamma} + b$, obtaining $a_{up} = 1.95$ and $b_{up} =-1.79 \cdot 10^{-4}$ ($R^2=0.97$) for the set of regions with $\rho_{\alpha}>1$, and $a_{down} =1.84$ and $b_{down} =-8.37 \cdot 10^{-4}$ ($R^2=0.99$) for the set of regions with $\rho_{\alpha}<1$; notice that $a_{up}/a_{down} \approx 1.1$. Binned data for the former set (triangles) and for the latter set (square) are shown in the main panel, together with the related best fits, in a log-log scale plot.
These fits are not significantly better that the one obtained at the country level, suggesting that possible internal heterogeneities may be  rather limited. In the insets we compare these best fits with raw data for two regions (Andalucia and Comunitat Valenciana) with $\rho_{\alpha}>1$ (upper inset) and two regions (Castilla y Leon and Cataluna) with $\rho_{\alpha}< 1$ (lower inset). Notice that in both cases data points overlap both curves, again suggesting that the division highlighted here is rather mild.}
\label{fig:ESnat}
\end{center}
\end{figure}

\begin{figure}[htb!]
\begin{center}
\includegraphics[scale=0.6]{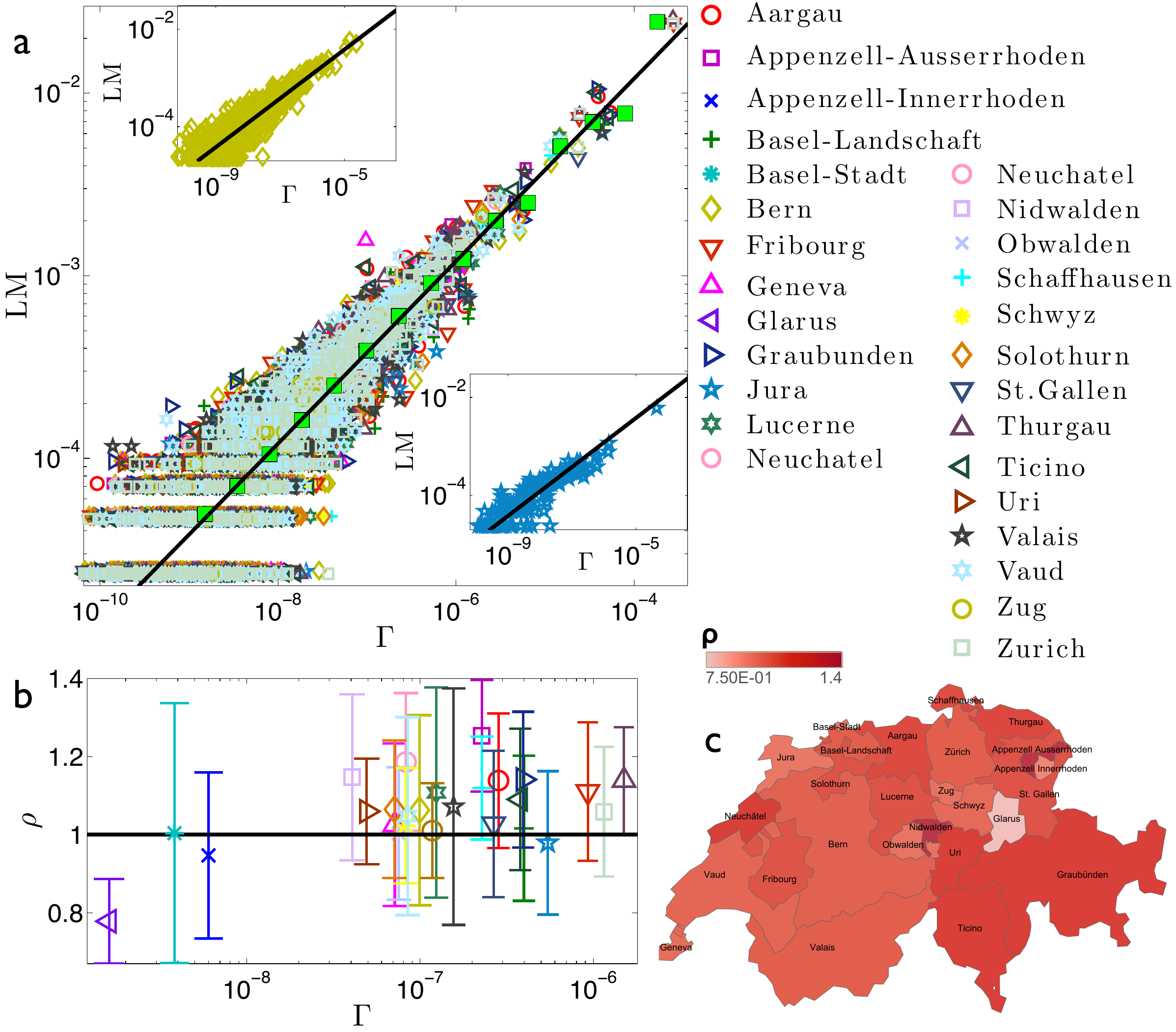}
\caption{Refined data analysis for local marriages in Switzerland. Panel $a$: log-log scale plot of $\textrm{LM}_{\alpha_i,y}$ versus $\Gamma_{\alpha_i,y}$, where different regions are denoted in different colours and symbols, as explained in the legend.
These data are properly binned (green squares) and best-fitted (solid line) by $y(\Gamma) = a \sqrt{\Gamma} + b$, in agreement with the theoretical result (\ref{scaling}). The best-fit coefficients are $a=1.20$ and $b=1.93 \cdot 10^{-7}$ with $R^2=0.99$.
Notice that the parameter $b$ accounts for the standard deviation associated to binned data. 
In general, the various regions seem to be homogeneously scattered around the best-fit curve.
In the insets we show, as examples, the data pertaining to two particular regions, namely ... (upper inset) and ... (lower inset). COMPLETA Notice that for both the best-fit previously obtained for the whole data set (solid line) still provides a very good fit.
Panel $b$: For each region we calculate $\rho_{\alpha}$, as defined in the text and deepened in Sec.~\ref{metodo-teorico}. The horizontal line is drawn as a reference for the unitary value. Notice that for most regions the deviation with respect the unitary value is within the error bar. The largest deviation from the unitary value is for Glarus, which also exhibits the smallest mean value of $\Gamma$. Given that, a few cases apart, all the regions are compatible with the overall best fit no further analysis on territorial homogeneity are performed.  The colormap presented in panel $c$ highlights the deviation of $\rho_{\alpha}$ with respect to the unitary value}
\label{fig:CHnat}
\end{center}
\end{figure}

Our aim is now to detect the possible existence of clusters of regions displaying different behaviours by comparing regional outcomes with the national average.
The analysis performed is summarized in Fig.~\ref{fig:FRnat} for France, Fig.~\ref{fig:DEnat} for Germany, Fig.~\ref{fig:ITnat} for Italy, Fig.~\ref{fig:ESnat} for Spain, and Fig.~\ref{fig:CHnat} for Switzerland. First, we look at how data pertaining to different regions are scattered around the best-fit obtained over the whole set of data (panels $a$). This analysis is able to immediately highlight large and systematic deviations of a subset of empirical data (e.g., pertaining to a particular region) with respect to the expected behaviour represented by the best-fit $f(\Gamma)$. Such deviations can further be quantified as follows. For each data point, say $(\Gamma_{\alpha_i, y}, \textrm{LM}_{\alpha_i, y})$, we calculate $\rho_{\alpha_i, y} = \textrm{LM}_{\alpha_i, y} / f(\Gamma_{\alpha_i, y})$, in such a way that a value of $\rho_{\alpha_i, y}$ close to (far from) one means that the best-fit provides a good (poor) estimate for the number of marriages in the province $\alpha_i$ at the year $y$. Further, we calculate the spatial and temporal average of $\rho_{\alpha_i, y}$, namely, we get $\rho_{\alpha} = \sum_{i=1}^{N_{P_{\alpha}}} \sum_{y=1}^T \rho_{\alpha_i, y}/(T \times N_{P_{\alpha}})$ and this is related to $\Gamma_{\alpha} = \sum_{i=1}^{N_{P_{\alpha}}} \sum_{y=1}^T  \Gamma_{\alpha_i, y}/(T \times N_{P_{\alpha}}) $ (panels $b$).
In this way we can inspect the possible presence of internal heterogeneity.
For instance, in the case of Switzerland, for any region (a couple of cases apart) $\rho_{\alpha}$ is, within the error bar, approximately unitary. This suggests that the average behaviour provided by the best fit constitutes a good representation for all the regions making up the country. A similar outcome emerges for Germany. Here, only two city-states (i.e., Berlin and Hamburg), both corresponding to relatively large densities of potential couples, fall significantly far from the average value.
On the other hand, for the remaining countries (i.e., France, Italy, and Spain), most regions exhibit a value of $\rho_{\alpha}$, which, still considering the related error, is not unitary.

For those countries we extend the analysis further and we distinguish between regions where the number of marriages is, respectively, underestimated (i.e., $\rho_{\alpha}>1$), overestimated (i.e., $\rho_{\alpha}<1$), and finely estimated (i.e., $\rho_{\alpha} \approx 1$) by the best fit. These cases are reported on the chart (panels $c$) with a colormap mirroring the value of $\rho_{\alpha}$: the brighter the color the smaller the related $\rho_{\alpha}$. It is immediate to see that regions corresponding to analogous outcomes tend to clusterize geographically. For instance, in France, the North-Eastern part and the Southern part form two distinct blocks. In Spain, the Southern part and the Basque region (including some adjacent regions) display analogous outcomes opposed to the West-most part. Finally, in Italy, the discrepancy between the two blocks is most evident and places side by side Northern and Southern regions.


One can therefore treat these clusters as different entities and derive the best fit for each of them separately and independently (panels $d$). The two best fits corresponding to regions with $\rho_{\alpha}<1$ and with  $\rho_{\alpha}>1$, respectively, are truly shifted only for Italy. This suggests that the heterogeneities emerging for France and Spain are weaker or, possibly, of different nature than those pertaining to Italy.

\subsection{Analysis of mixed marriages}

We now analyze, in a similar fashion, marriages between native and foreign-born citizens. The theoretical framework underlying this kind of phenomenology is analogous to the one used above, but here the two parties are played by native and by foreign-born individuals (rather than males and females, despite the constraint of heterogamy clearly connects these sets of variables as explained in section Methods); we refer to \cite{barra-SR,agliari-NJP} for a full treatment.

\begin{figure}[htb!]
\begin{center}
\includegraphics[scale=0.38]{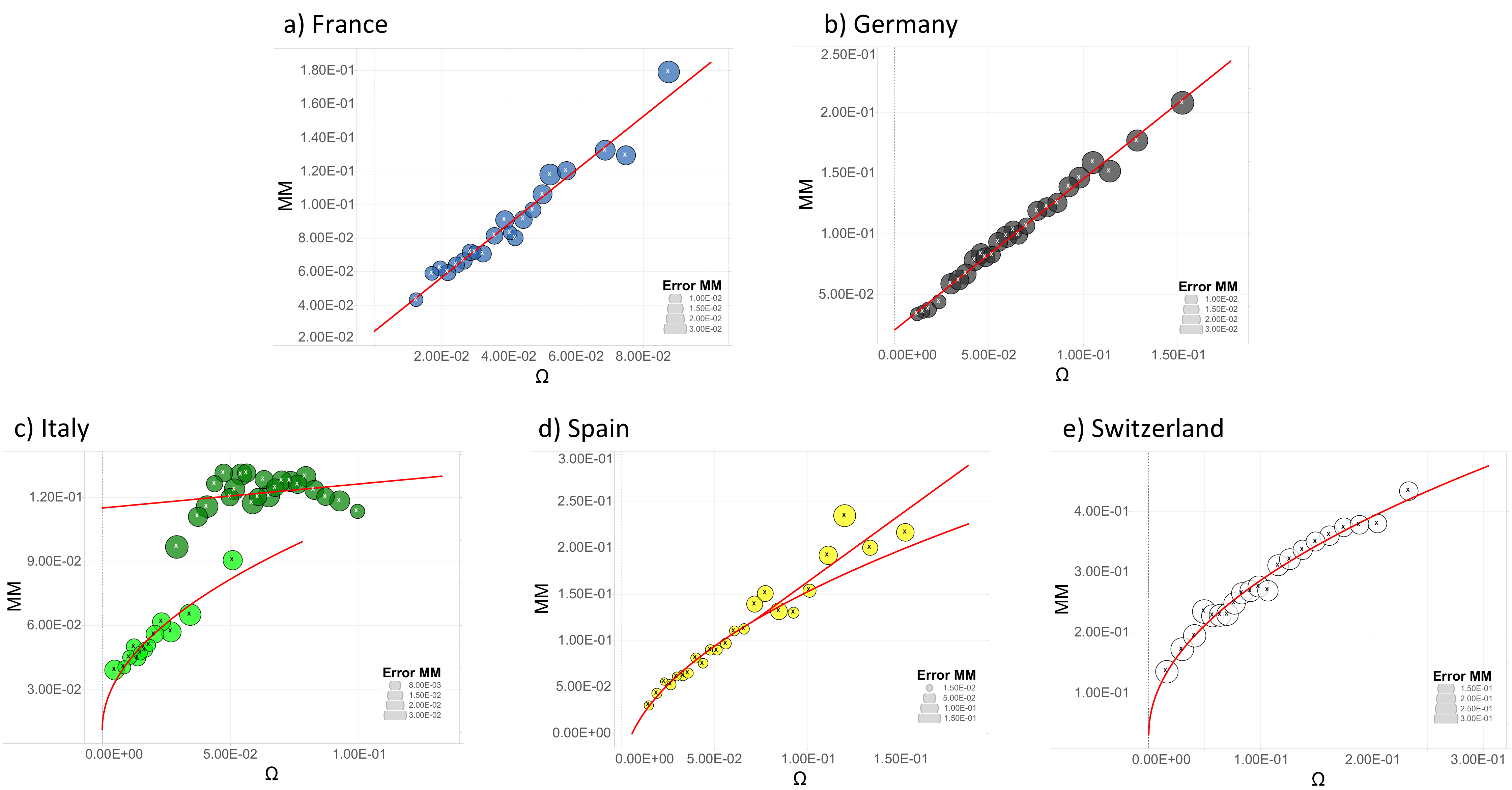}
\caption{Mixed marriages MM (that is involving a native citizen, either male or female, with an immigrant) versus the fraction of potential couples $\Omega$. Here the raw data used to built the points are $\textrm{MM}_{i,y}$, which is the number of marriages between native citizens and immigrants in the province $i$ at time $y$ divided by the total number of marriages (defined as before) in the province $i$ at time $y$, and $\Gamma_{i,y}$, that is the number of mixed couples (here the number of native citizens times the number of immigrants) in the province $i$ at time $y$ divided by the square of the sum of the two group of people in the province $i$ at time $y$. The error circles should be understood with respect to the legend reported in each graph. To optimize the visualization of the trends, the plots were realized dividing the interval of data of each state in bins of the following numbers (we write the name of the country followed by the amount of bins used): Germany 25, France 21, Italy 35, Switzerland 21, Spain 27. The red lines in the plots are the best fit functions. With respect to Fig. 1, here the different countries do not show the same behaviour: France and Germany follow a linear behaviour, which is derived from a non-interacting particles approach, while Spain and Switzerland maintain a square-root trend. Italy displays a mixed behaviour which highlights even more the difference between the north and south of the country: in fact, the north is best fitted by a square root function while the south by a line.}
\label{fig:misti}
\end{center}
\end{figure}

We check the theoretical law given by Eqs.~\ref{scalingM} and \ref{scaling} versus the available experimental data for the above mentioned five countries and we report our findings in Figure \ref{fig:misti}.
\newline
First, we notice that different countries display qualitatively different behaviours: the growth of mixed marriages is well described by a square-root law (at least for small values of $\Omega$) in Spain and in Switzerland, while in France and in Germany it is better described by a linear law; in Italy the behaviour is even more complex as data are rather noisy and two distinct trends emerge. Therefore, imitative interactions among native and foreign-born agents still seem to play a major role in Spain and in Switzerland, while in France and in Germany the presence of ``external fields'' seem to prevail. In Italy, the two trends can be associated to Northern and Southern regions, hence confirming strong discrepancies between the two geographical areas also as for immigrant integration.
\newline
Moreover, in general, the goodness of the fits is lower than the case of autochthonous marriages, although the size of the available data sets are the same. This suggests that a certain degree of non-uniformity may affect the data considered; for instance, no clusterization of data based on e.g., the provenance or the status of the foreign-born spouse is possible due to a lack of information. We also expect that the existence of large, well-established ethnical group may influence the degree of integration (in terms of mixed marriages) of individuals belonging to the group itself, but a complete quantification of this correlation is currently out of reach. These features may by crucial in future outlooks for getting a robust and sound picture of the phenomenology. For instance, here we just notice that most immigrants in Switzerland come from European (EU-28/EFTA) countries (in particular, $\approx 40\%$ of the immigrants come from the neighboring European countries and only $\approx 12 \%$ come from America and Africa) and
the number of foreign citizens living in the country is almost one fourth of the permanent resident population \cite{FSO}. In Germany, the fraction of foreign-born individuals is $\approx 12 \%$ and, being the second most popular migration destination in the world (after the United States), it attracts a wide-ranging class of immigrants (from refugees to high-professional figures) \cite{DESTATIS}. The flow of immigrants is very large in France as well with $\approx 11 \%$ of foreign born individuals, of which $\approx 32 \%$ come from Europe and $\approx 43 \%$ come from Africa (mostly from French-speaking countries) \cite{INSEE}. 
Immigration from (former) colonies is significant also in Spain, where $\approx 25\%$ of the immigrants come from South and Central America, $\approx 41 \%$ come from European (EU-28/EFTA) countries and $\approx 18 \%$ from Africa \cite{INE}. As for Italy, almost one fourth of the foreign-born population comes from Romania, another fourth coming from Albania, Morocco and China together \cite{ISTAT}.

\section{Methods}\label{metodo-teorico}
In this section we deepen the Statistical Mechanics approach underlying our analysis and leading to Eqs.~\ref{scalingM} and \ref{scaling}, as well as the methodology followed during data analysis and leading to the empirical evidence described above.

\subsection{The theoretical protocol}

The theoretical modeling is split in two parts, each covering one of the two models used as reference guide. We first discuss the one-body theory (independent model) and then move to the two-body theory (model with social interactions).
\newline
In both cases we describe the system in terms of decision makers, whose ``decision'' is denoted with $\sigma_{\mu}$, $\mu=1,...,M_i$ and with $\tau_{\nu}$, $\nu=1,...,F_i$, where $\mu$ and $\nu$ label, respectively, the generic $\mu^{th}$ male and the generic $\nu^{th}$ female within the $i$ province, and $M_i$ and $F_i$ represents the total male and female populations within the province $i$ (according to the case considered the population will be restricted to natives or to natives and immigrants). 
More precisely, $\sigma_{\mu}$ and $\tau_{\nu}$ are taken binary and meant to denote the attitude toward marriage: $\sigma_{\mu} = +1$ and $\tau_{\nu}=+1$ indicate a positive bias to contract marriage and, vice versa, $\sigma_{\mu} = -1$ and $\tau_{\nu}=-1$ indicate a negative bias to contract marriage. These decisions can be modulated by external influences (Mc Fadden scenario) or by peer interactions (Brock and Durlauf scenario): in both cases, these ingredients are added in terms of a cost function $H(\{\sigma\}, \{\tau\})$ (i.e an Hamiltonian) to be minimized; for the former the Hamiltonian is {\em one-body}, that is, it contains no interactions between variables $\sigma, \tau$,  i.e. $H^{(1)}(\{\sigma\}, \{\tau\})\sim - h ( \sum_{\mu} \sigma_{\mu} + \sum_{\nu} \tau_{\nu})$, while for the latter the Hamiltonian contains interactions among agents, i.e. $H^{(2)}(\{\sigma\}, \{\tau\}) \sim - \sum_{\mu, \nu} \sigma_{\mu} \tau_{\nu}$. Whatever the choice, within this approach, minimization of the cost function is coupled with the request to simultaneously maximization of the related resulting entropy. This request, often called Maximum Entropy Approach in inferential investigations  \cite{bialek,cavagna}, is equivalent to assuming the less structured model, whose predictions (e.g. first and second moments) are in agreement with the experimental data. Statistical Mechanics has a deep dual structure within the classical statistical routes as pointed out by Jaynes \cite{yanes1,yanes2}  and returns the probabilistic weight for a given configuration $\{ \sigma, \tau \}$ in terms of the Maxwell-Boltzman factor $P(\sigma, \tau) \propto \exp[-H(\{\sigma\}, \{\tau\})]$.
\newline
This inferential estimation for mean-field imitative models has been recently extensively treated in \cite{fedele}, where the interested reader may further deepen the foundation of this approach.

\subsubsection{Independent model (for mixed marriages)}

In this subsection we discuss the one-body theory (referring to \cite{agliari-NJP,barra-SR} for extensive details). Since the outcomes of this theory have been shown to be in agreement only to the case of mixed marriages (see Sec.~\ref{sec:results}), we develop the mathematical scaffold already thinking at mixed marriages (i.e. $MM$) and in terms of possible mixed couples (i.e. $\Omega$). In the {\em independent model}, reciprocal influences among decision makers are not allowed for and the expected number of mixed marriages can be estimated by a simple probabilistic approach, where only the relative sizes matter. Since the (expected) universal linear scaling between the amount of males and females within each province is always respected (both for natives and for migrants), we can write (directly at the level of the whole nation, see Fig. \ref{fig:popolazione} and Sec.~\ref{metodo-sperimentale})
\begin{equation}
MM = h\  \omega (1-\omega) \equiv h\ \Omega,
\end{equation}
where $h$ represent the overall realizability of a mixed marriage (characterizing the couple guest/host country) and $\omega (1-\omega)$ is the probability of a mixed couple meeting, drawn indipendently from a box of $\omega$ (fraction of natives) and $1-\omega$ (fraction of foreign-borns) citizens.

This behaviour can be encoded as well with a one-body cost function depending on an external field only:  
calling $\sigma_{\mu}^{i}= \pm 1$, $(\mu=1,...,M_{i})$, the positive ($\sigma=+1$) -or negative ($\sigma=-1$)- attitude of the $\mu^{th}$ male agent within the province $i$ to contract the marriage, calling $\tau_{\nu}^{i}= \pm1$, $(\nu=1,...,F_{i})$ to account for the will of female decision makers and, finally, introducing also $h\Omega$ as properly {\em field} containing the probability of a mixed encounter, the model is coded into
\be \label{1body}
H^{(1)}(\{\sigma\}, \{\tau\}) = - \sum_{i=1}^{N_P} h\Omega \left(\sum_{\mu=1}^{M_i}\sigma^i_{\mu} + \sum_{\nu=1}^{F_i} \tau^i_{\nu}\right)
\ee
thus energy minimization simply suggests that the agents on average try to behave accordingly to the external suggestions encoded in the fields $h$, that is, positive values of $h$ favor marriages and viceversa (and the larger the value, the stronger the effect).
Solving one-body models is straightforward in Statistical Mechanics and returns a trend for mixed marriages as
\be \label{lin}
MM \propto \frac{1}{N_P}\sum_{i=1}^{N_P}\frac{1}{M_i+F_i} \left( \sum_{\mu=1}^{M_i} \left\langle \sigma_{\mu}^i\right\rangle  + \sum_{\nu=1}^{F_i} \left\langle \tau_{\nu}^i \right\rangle \right)\sim h \Omega,
\ee
namely marriages driven mainly by external influences (over peer interactions) are expected to scale linearly in the volume of possible couples.

\subsubsection{Model with social interactions (for local marriages)}

In this subsection in turn we discuss the two-body theory (referring to \cite{agliari-NJP,barra-SR} for extensive details): as outcomes of this theory apply to all the local marriages, we develop the mathematical scaffold already thinking at local marriages (i.e., LM) and in terms of possible autochthonous couples (i.e. $\Gamma$).
\newline
In order to develop the theory leading to Eq.~\ref{scaling}, we need to introduce a (two-body) {\em Hamiltonian} $H^{(2)}(\{\sigma\}, \{\tau\})$, describing the interactions among males and females \footnote{We considered only heterosexual couples as, over the time window considered, homo-sexual marriages where still forbidden in the countries considered.}. Each individual $\mu, \nu$ is associated to a dichotomic variable or ``spin'' (referred to as $\sigma_{\mu}=\pm 1$ for males and as $\tau_{\nu}=\pm 1$ for females) encoding a positive (i.e. $+1$) or negative (i.e. $-1$) attitude to marriage. Then, the ``cost'' of a given configuration $(\{\sigma\}, \{\tau\})$ is provided by the global Hamiltonian
\be\label{Hamiltoniana}
H^{(2)}(\{\sigma\}, \{\tau\}) = - \frac{J}{N}\sum_{i=1}^{N_P} \left( \sum_{\mu=1}^{M_{i}} \sum_{\nu=1}^{F_{i}} (\sigma_{\mu}^{i}\tau_{\nu}^{i})\right),
\ee
where $N = \sum_{i} (M_{i} + F_{i})$ is the total population in the whole country and $J$ tuning the interaction strenght.
%
%
%
%
This Hamiltonian is simply the sum of terms like $-\sigma^i_{\mu} \tau^i_{\nu}$ over all the possible couples $(\mu,\nu)$ of individuals belonging to the same province $i$. Clearly, the underlying {\em mean field} assumption of a fully connected network is only a working approximation as real social networks are expected to be small worlds \cite{Agliari-grano,granovetter1,granovetter2}, however, for simple enough interaction rules (as the imitative one coded in eq. (\ref{Hamiltoniana}) in the present work) the general scenario (i.e. criticality, scaling, etc.) provided by the mean field approximation suitably approximates the more realistic one obtained embedding the system on a small world network \cite{barra-PhysicaA,watts,sollich,martin}.

Intuitively, for the minimum energy principle -that tries to keep the numerical value of $H(\{\sigma\}, \{\tau\})$ at its minimum \cite{ellis}- the function (\ref{Hamiltoniana})  favors the configurations of citizens with the overall lowest possible frustration, that is, considering the couple $(\mu,\nu)$ as an example, it favors the two states where the variables $\sigma_{\mu}$ and $\tau_{\nu}$ are aligned, thus $\sigma_{\mu}=\tau_{\nu}=+1$ or $\sigma_{\mu}=\tau_{\nu}=-1$, while misaligned couples $\sigma_{\mu} \neq \tau_{\nu}$ are unfavored. This captures the trivial observation that, within any country,  stable couples ($\sigma_{\mu} = \tau_{\nu} = +1$) as well as
stable {\em not-couples} ($\sigma_{\mu} = \tau_{\nu} = -1$) exist
\footnote{The existence of stable not-couples may be less intuitive yet these are largely predominant. To be convinced regarding the existence and abundance of stable {\em not-couples} it is already enough to note that, within a given country, every single decision maker is in contact with a tiny fraction only of the whole set of decision makers of the opposite sex present in the territory.}. On the other hand, the conflicting situation where one of the two partners wants to get married (say $\sigma_{\mu}=+1$) but the other does not (hence $\tau_{\mu}=-1$) is only  transitory (hence not stable) as, after a proper timescale, the pretender is expected to move toward another target (pathological cases apart, whose presence may act as a small noise over the mean results).
\newline
Once the Hamiltonian ruling the phenomenon is assigned, this allows introducing in the standard way the partition function $Z$ related to the present case (and whose explicit expression permits to obtain all the desired information) as
\be
Z = \sum_{\{\sigma\}} \sum_{\{\tau\}}\exp\left[ -  H( \{\sigma\}, \{\tau\})  \right],
\ee
where the sum is performed over all possible $\prod_{i=1}^{N_P}2^{M_{i}} \times 2^{F_{i}}$ configurations.
By a direct calculation, it is straightforward to check that
\be
\label{uno}
Z_{N} = \sum_{\{\sigma\}} \sum_{\{\tau\}}\exp\left[ \frac{J}{N}\sum_{i=1}^{N_P}\left( \sum_{\mu=1}^{M_{i}} \sum_{\nu=1}^{F_{i}} \sigma_{\mu}^{i}\tau_{\nu}^{i}\right) \right]
\sim \sum_{\{\sigma\}} \exp\left \{ \sum_{i=1}^{N_P}  \left[ \frac12 M_{i} J^2 \Gamma_{i} m^2_{i} \right] \right \},
\ee
where $\Gamma_i=M_iF_i/N$ and we highlighted the term $m_{i} =  (\sum_{\mu}^{M_{i}}\sigma_{\mu}) / M_{i}$, that measures the average propensity to get married for males within the province $i$. This quantity works as the ``order parameter'' of the theory and it is expected to be proportional to the experimental estimate of the amount of marriages $\textrm{LM}_{i}$ within the province $i$ (suitably normalized). Note that, in order to study the joint evolution of male and female attitudes to marriage, the requirement of heterosexual marriages implicitly allows to study  the average propensity of only one of the two parties (here the male one, via $m_{i}$, while clearly the same results are achievable using female magnetization as the order parameter).

Before proceeding it is worth stressing the {\em social meaning} of the last and crucial passage in eq. (\ref{uno})\footnote{Note that the equivalence is actually rather robust as it holds in the presence of general positive couplings between $\sigma$ and $\tau$, even in inhomogeneous and/or diluted networks \cite{agliari-EPL}, and for binary as well as real (i.e. Gaussian) variables \cite{barra-SR}}: this equivalence is trivially obtained by performing the summation over the $\tau$ variables (i.e., by integrating over the ``female degrees of freedom''), that returns a term $\cosh [J(M_{i}/N) m_{i}]^{F_i}$; the latter  is then written using $\cosh(x) = \exp [\ln\cosh(x)]$ and then Taylor-expanding at the leading term as $\ln\cosh(x)\sim x^2/2$.
\newline
Remarkably, such an equivalence states that the initial model described by eq. (\ref{Hamiltoniana}), meant for males and females in interaction and encoded by sums of terms $\propto  - \sigma_{\mu}^{i}\tau_{\nu}^{i}$, is statistically equivalent to a model accounting for imitative interactions among males only, thus  encoded by terms $\propto - \sigma_{\mu}^{i}\sigma_{\nu}^{i}$, that is $- m_{i}^2$. Otherwise stated, the phenomenological rule described by the cost function (\ref{Hamiltoniana}), where males and females interact trying to satisfy their relationships, recovers the copy model theory, that is, the discrete choice  with imitation \cite{brock,cont} in socio-economic literature (or Hebbian ferromagnetism in statistical mechanics literature \cite{agliari-EPL,barra-JSP,barra-PhysicaA,ginestra}).
Of course, and in complete analogy, we could reach imitation among females only, by summing over the $\sigma$ variables first.
Discrete-choices with imitation, where agents (here males) interact pair-wise in an imitative fashion, is well known in statistical mechanics (as the Curie-Weiss model \cite{barra-JSP,pizzo}) as well as in quantitative sociology (as the Brock-Daurlauf theory \cite{brock3}): for this model, where interactions between citizens  from different provinces were neglected, the expected (i.e. averaged) behavior of the order parameter $m_i$ depends only on the parameter $\Gamma_i$ in the form
\be
m = \tanh \left[ J^2 \Gamma \cdot m\right],
\ee
By Taylor-expanding the above equation for small $m$ we get $m \propto \sqrt{\Gamma}$ and identifying the theoretical order parameter $m$ with the experimental one LM, we obtain the expected leading trend
\be\label{square}
\textrm{LM} \sim \sqrt {  \Gamma },
\ee
This equation predicts, within each country (namely under the assumption of homogeneity among the $\sigma$ as well as the $\tau$ variables), a square-root growth for the expected number of marriages LM versus the density of potential couples $\Gamma$ and can  be compared directly with experimental data.

\subsection{The experimental protocol}\label{metodo-sperimentale}

Even the experimental protocol is split into two main parts. In the first one we revise how we elaborated the time series to recover observables to be framed in the Statistical Mechanics model (i.e. the results presented in Fig.~\ref{fig:local}), that is also the route paved in \cite{agliari-NJP,barra-SR}; then, in the second section we discuss novel inferential techniques we developed to reveal potential regional clusters (used to produce Figures \ref{fig:FRnat}-\ref{fig:CHnat}).

\subsubsection{General scenario}

\begin{figure}[H]
\centering
\includegraphics[scale=0.34]{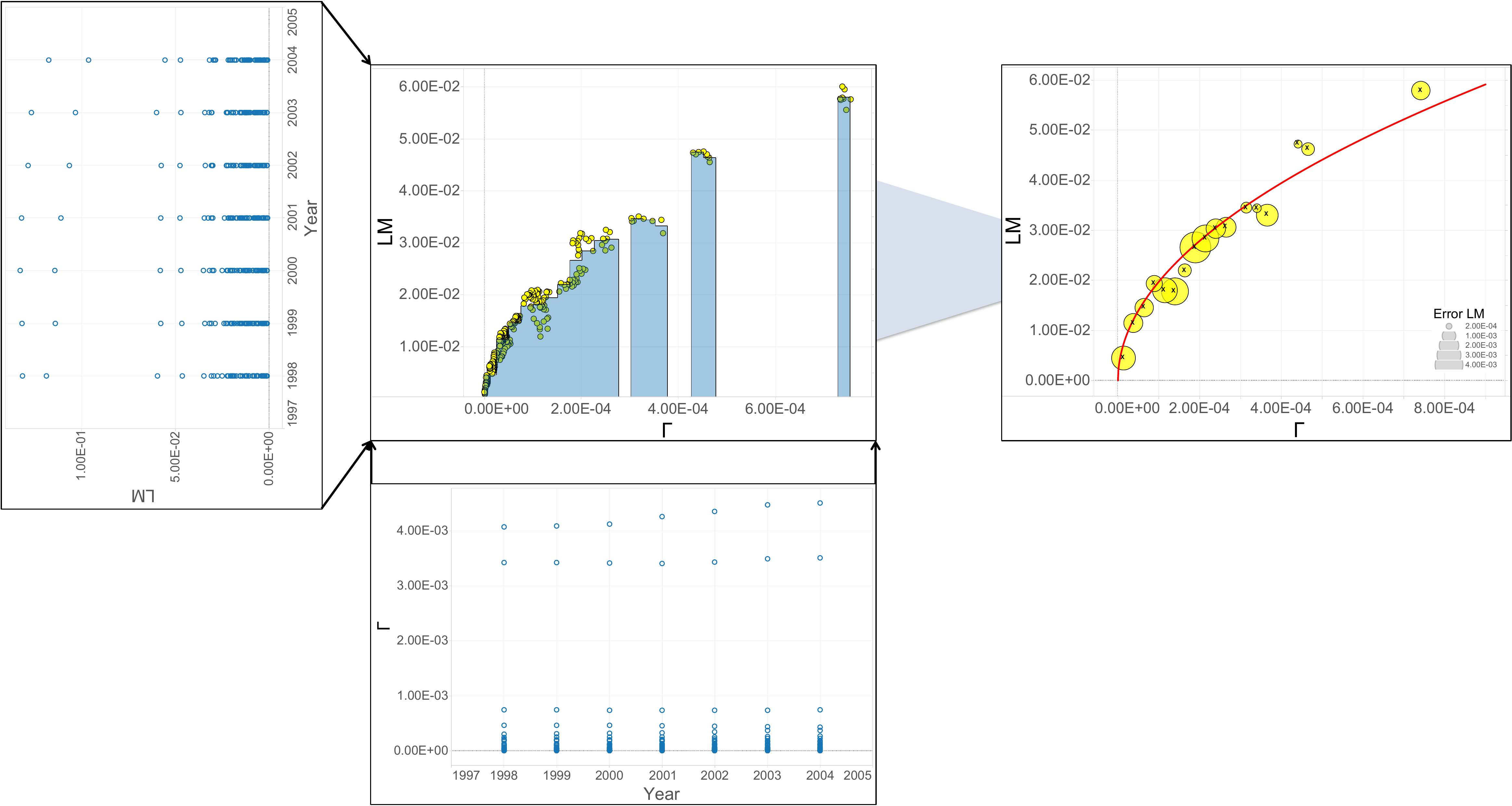}
\caption{Schematic representation of the procedure performed in data analysis. For each country (here we show only local marriages in Spain as test case) the time series for fraction of marriages and fraction of couples are considered, as shown in the left panel. Each dot corresponds to a given province and a given year. We check that, for each province $i$, $\Gamma_{i,y}$ (or $\Omega$ according to the case considered) is monotonic versus time, so to allow the inversion that returns time versus $\Gamma$ (not shown). Then, time dependence is by-passed by plotting -as raw data- the fraction of marriages versus the fraction of couples as shown in the central panel. Lastly we bin the latter to obtain the coarse grained evolution of the phenomenon, whose data points (yellow circles) have been best fitted against the theory (solid line).}
\label{fig:data}
\end{figure}

We collected data for the number of marriages (local and mixed) and for the female and male populations (native and foreign-born) in France (source: INSEE), Germany (source: DESTATIS), Italy (source: ISTAT), Spain (source: INE), and Switzerland (source: FSO) over the time windows $[2006-2010]$, $[2007-2012]$, $[2005-2010]$, $[1996,1998-2004]$\footnote{In the case of mixed marriages in Spain it is $T= 6$ [1999-2004]}, $[2008-2010]$, respectively.
Time series are collected at the provincial level. This degree of resolution is determined by the empirical observation that marriages typically involve people living in the same province (i.e., the social interactions display a characteristic geographic scale \cite{agliari-NJP}); further, this allows, in turn, treating each province independently from the others and thus performing statistical analysis over the set of provinces meant as different realisations of the same system.

Therefore, for any given country, we have $N_P$ provinces and for each, labeled as $i$, we collected the data series $\{ \textrm{LM}_{i,y}\}$, $\{ \textrm{MM}_{i,y}\}$, $\{\Gamma_{i,y}\}$, and $\{\Omega_{i,y}\}$,  where $y$ is a time index, running over the time window considered (we recall that data are sampled yearly). More precisely,\\
$\bullet$ $\textrm{LM}_{i,y}$ represents the ratio between the number of \emph{local} marriages (i.e., between two autochthonous people) registered in the province $i$ at time $y$
and the number of \emph{all} marriages (i.e., including local marriages, mixed marriages and marriages between two foreign-born people) registered at time $y$;\\
$\bullet$ $\textrm{MM}_{i,y}$ represents the ratio between the number of \emph{mixed} marriages (i.e., between one native and one foreign-born) registered in the province $i$ at time $y$ and the number of \emph{all} marriages (i.e., including local marriages, mixed marriages and marriages between two foreign-born people) registered in the province $i$ at time $y$;\\
$\bullet$ $\Gamma_{i,y}$ represents the number of possible couples in the province $i$ at time $y$ and is defined as the the number of males times the number of females (residing in the province $i$ at time $y$) divided by the square of the overall population (residing in the country at time $y$);\\
$\bullet$ $\Omega_{i,y} $ represents the number of possible mixed couples in the province $i$ at time $y$ and is defined as the the number of natives times the number of foreign-born (residing in the province $i$ at time $y$) divided by the square of the overall population (residing in the province $i$ at time $y$).

\begin{figure}[htb!]
\begin{center}
\includegraphics[scale=0.4]{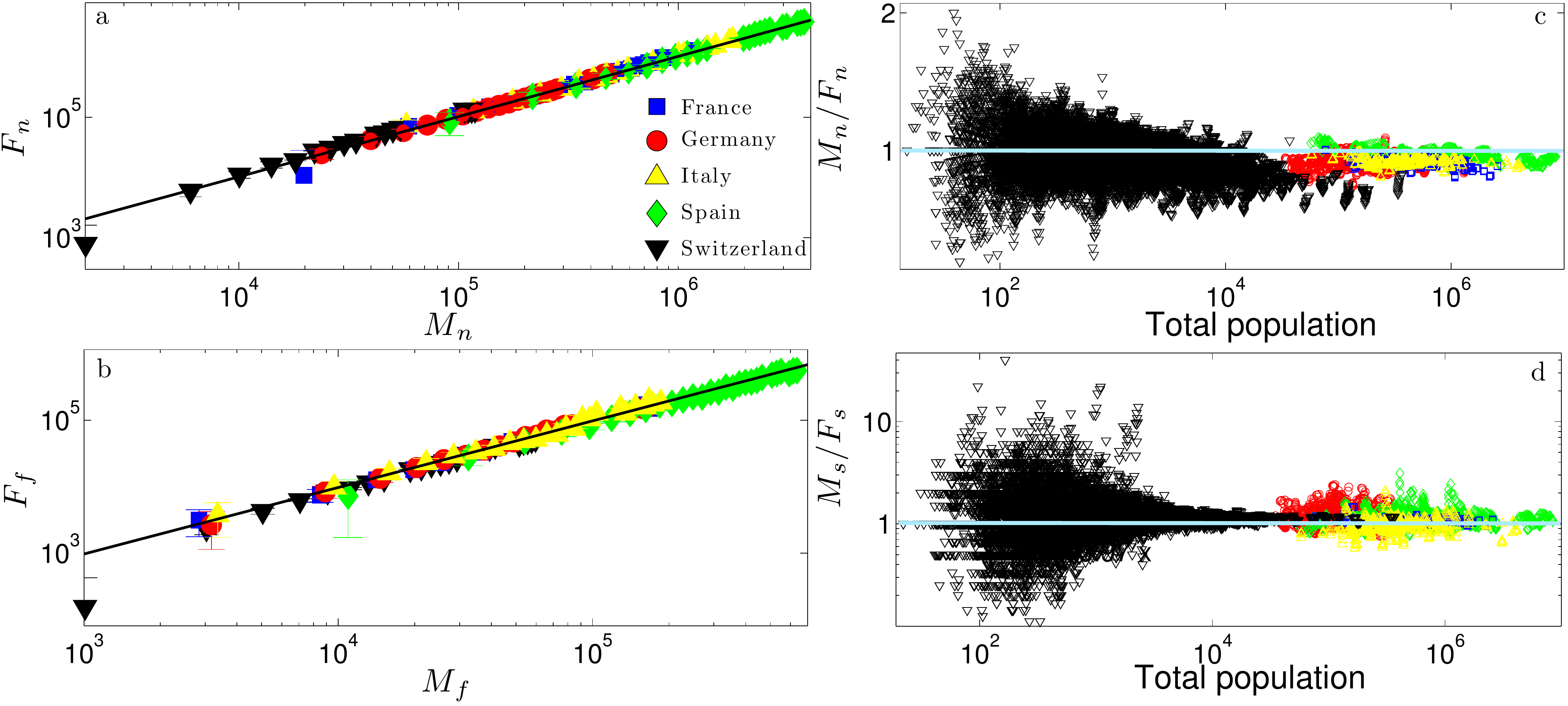} \\
\caption{Panel $a$: log-log scale plot for the number of native females versus the number of native males. Binned data for different countries are represented in different symbols and colours as explained in the legend; the same legend holds for panels $b$, $c$, and $d$ as well. The solid line represents the best fit given by $y = a x + b$, where $a \approx 1.03$. Panel $b$:  log-log scale plot for number of foreign-born females versus the number of foreign-born males; each point represent the average value over the whole country and for a different year. The solid line represents the best fit given by $y = a x + b$, where $a \approx 0.97$. Panel $c$: scatter plot for the ratio between the number of native males and the number of native females. Each data point represents the value obtained for a different year and province. The solid line represents the unitary value as a reference. Panel $d$: scatter plot for the ratio between the number of foreign-born males and the number of foreign-born females. Each data point represents the value obtained for a different year and province. The solid line represents the unitary value as a reference.}
\label{fig:popolazione}
\end{center}
\end{figure}

We stress that data on marriages account only for heterosexual marriages and that $\Gamma_{i,y}$ accordingly refers to heterosexual couples. On the other hand, by definition, $\Omega_{i,y}$ includes all possible pairs. This choice is meant to simplify the treatment, since  in the Statistical Mechanics analysis one has to deal with a bipartite model rather than a system made of four parties. This simplifications is justified by the following empirical observation (see Fig.~\ref{fig:popolazione}): the immigrant community is approximately made of half males and half females and this ratio is independent of the extent of the community. By the way, the same is evidenced also for the native community as shown in Fig.~\ref{fig:popolazione} too.
\newline
As a result, denoting with $F_f$ the number of female foreign-born agents, with $M_f$ the number of male foreign-born agents, with $F_n$ the number of female native agents, and with $M_n$ the number of male foreign-born, one has
$F_f \approx M_f \approx (F_f + M_f )/2$ and $F_n \approx M_n \approx (F_n + M_n )/2$. Also, $\Omega \approx (F_f + M_f)/(F_f + M_f + F_n + M_n )$.
The fraction of possible heterosexual mixed couples can therefore be written as
\begin{eqnarray}\label{cinque}
    \frac{M_n \cdot F_f +  M_f \cdot F_n}{(M_n+M_f)\cdot(F_n+F_F)}&=& 2 \omega (1-\omega) = 2 \Omega.
\end{eqnarray}
Thus, it is reasonably to take $\Omega$ as an estimate for the number of possible mixed couples.


The time series described above are then manipulated according to a lengthy but simple protocol, largely discussed in \cite{barra-SR,agliari-NJP} and summarized in Figure \ref{fig:data}, leading to a set of binned data points for local and mixed marriages versus $\Gamma$ and $\Omega$, respectively. The dependence on the province $i$ is lost during manipulation and under the hypothesis that provinces making up the same country are sufficiently homogeneous (this point is further deepened below). The dependence on $y$ is lost as well or, otherwise stated, it accounts for the variation of $\Gamma$ and $\Omega$.

In this way we extract the average evolution of LM versus $\Gamma$ and of MM versus $\Omega$, and these are then best-fitted against the laws stemming from the theoretical analyses, namely the square-root law Eq.~\ref{square} and the linear law Eq.~\ref{lin}.

Results for local and mixed marriages are shown in Figs.~\ref{fig:local} and \ref{fig:misti}, respectively, and are discussed in Sec.~\ref{sec:results}.

\subsubsection{Cluster detection}

Once the overall behavior at the country level has been inferred, there is still to face the homogeneity issue, that is, the existence of regions displaying systematic differences with respect to the average results. The analysis is meant to highlight the potential presence of inner differences, possible heritages of ancient different cultural traditions, thus it is performed only on local marriages.

Figure \ref{fig:local} is obtained averaging over all the provinces within the country, as described in the previous subsection. The binned data points are then best-fitted versus the function $f(\Gamma) = a \sqrt{\Gamma} + b$ in agreement with the theoretical model (\ref{square}).
The next step is to refine the level of resolution and to distinguish different regions;  the behavior of each region is then compared to the average one (i.e. the one obtained at the country level).
Therefore, the above mentioned time series have been reshuffled as $\{\Gamma_{\alpha_i,y}\}$ and $\{ \textrm{LM}_{\alpha_i,y}\}$, where $\alpha_i$ denotes the $i$-th province in the $\alpha$-th region. Of course, each country displays a different number $N_R$ of regions ($N_R = 22$ for France, $N_R = 16$ for Germany, $N_R = 20$ for Italy, $N_R = 18$ for Spain, $N_R = 26$ for Switzerland) and each region $\alpha$ includes a different number $N_{P_{\alpha}}$ of provinces.

In Figures \ref{fig:FRnat}-\ref{fig:CHnat} the set of data $\{ \textrm{LM}_{\alpha_i,y}\}$ is plotted versus $\{\Gamma_{\alpha_i,y}\}$ and each region is depicted in a different color. In this way one can immediately detect anomalous behaviours with respect to the average behaviour represented by the function $y = f(\Gamma)$. For instance, with this method, we recover that Corsica (see \ref{fig:FRnat}) and Melila (see \ref{fig:ESnat}) do not match the reference curve, as somehow expected given the peculiarity of these regions. In order to quantify the goodness of the estimate provided by the best fit, we introduce the observable $\rho_{\alpha_i,y}$ and its average $\rho_{\alpha}$ defined, respectively, as
\begin{eqnarray}
\rho_{\alpha_i,y} = \frac{f(\Gamma_{y,\alpha})}{\textrm{LM}_{\alpha_i,y}},\\
\rho_{\alpha} = \frac{1}{N_P} \sum_{i=1}^{N_P} \rho_{\alpha_i}.
\end{eqnarray}

\begin{figure}[H]
\centering
\includegraphics[scale=0.42]{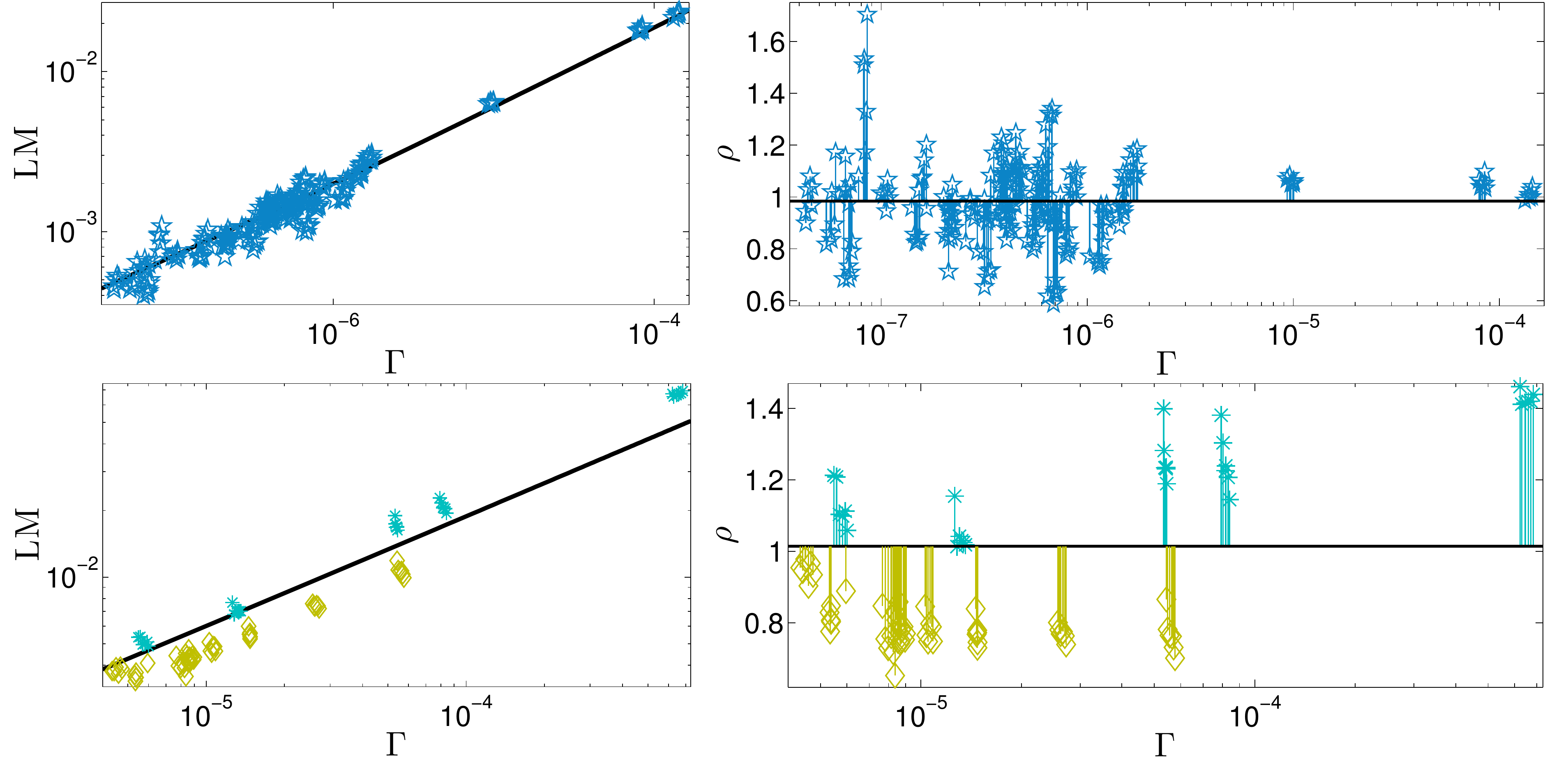}
\caption{The dispersion of the observable $\rho_{\alpha_i}$ is shown for the region Rheinland-Pfalz in Germany (upper panel) and for the regions Campania and Emilia-Romagna in Italy (lower panel). More precisely, in panel $a$ we show the set of data $\textrm{LM}_{\alpha_i}$ versus $\Gamma_{\alpha_i}$, where $\alpha_i$ here runs over all the $x$ provinces in Rheinland-Pfalz; the solid black line represents the best-fit obtained over the whole set of data pertaining to Germany. Notice that the best-fit is the same already presented in Fig.~\ref{fig:DEnat}. In panel $b$ we show $\rho_{\alpha_i}$ obtained by dividing each data point $\textrm{LM}_{\alpha_i}$ in panel $a$ by the related expected value $f(\Gamma_{\alpha_i})$. Notice that the values of $\rho_{\alpha_i}$ are scattered randomly around $1$. The same passages are applied, mutatis mutandis, in panels $c$ and $d$; the legend and the best fit are the same as in Fig.~\ref{fig:ITnat}. Notice that now data pertaining to Campania and to Emilia-Romagna are systematically above and below, respectively, the best-fit line. Accordingly, the related values of $\rho_{\alpha_i}$ are always larger and smaller than $1$, respectively.}
\label{fig:cluster}
\end{figure}

In a country where people behave homogeneously throughout its territory one expects that $\rho_{\alpha_i}$ fluctuates randomly around $1$, in such a way that its average $\rho_{\alpha}$ is close to one. This is the case for regions in Germany and in Switzerland, see Fig.~\ref{fig:cluster} (upper panels). On the other hand, in a country displaying internal heterogeneities, we expect that for some regions (possibly forming a connected cluster) $\rho_{\alpha_i}$ is systematically above (or below) $1$, in such a way that its average $\rho_{\alpha}$ is far from one.
This is the case for regions in France, Italy and Spain, see Fig.~\ref{fig:cluster} (lower panels).

Notice that $\rho_{\alpha}>1$ means that the global best-fit overestimates the number of marriages in the region $\alpha$, while $\rho_{\alpha}<1$ means that the global best-fit underestimates the number of marriages in the region $\alpha$. Therefore, the simplest distinction one can introduce is between regions where the number of marriages is relatively low (i.e., $\rho_{\alpha}>1$) and regions where the number of marriages is relatively high (i.e., $\rho_{\alpha}<1$).
This distinction is performed for France, Italy and Spain, as outlined in Figs.~\ref{fig:FRnat}, \ref{fig:ITnat}, \ref{fig:ESnat}. Interestingly, regions displaying the same trend constitute clusters.

Each cluster is then treated separately, and for each a new best-fit is found. The same analysis as before are repeated and again compared with data. A true shift is actually evidenced only for Italy.

\section{Conclusions} \label{sec:conclusions}

In this paper we analyzed, within a statistical mechanical framework, the way the number of marriages evolves in five different European countries; we consider local marriages (i.e. involving two autochthonous spouses) and mixed marriages (i.e. involving a native spouse and a foreign-born spouse). The inspected countries are France, Germany, Italy, Spain and Switzerland. Instead of classical historical series analysis (i.e. the study of temporal dynamics of the social quantifier considered), for each of these countries we studied the evolution in the density of marriages (local and mixed) versus the percentage of potential couples (both natives or mixed, respectively) and we find that the results can all be framed within the two extrema scenarios:
\newline
- Mc Fadden Discrete Choice: each agent decides regarding his/her will to get married essentially without relying on peer choices. Within this framework, elementary statistical mechanical calculations predict a linear correlation between the density of marriages and the density of potential couples present in the territory.
\newline
- Brock and Durlauf Imitational Choice: each agent decides according to imitational mechanisms based on peer choices. Within this framework, statistical mechanics of imitative models (i.e. ferromagnetic models) predicts a square root relation between the density of marriages and the density of potential couples present in the territory.

After a proper data manipulation and model calibration we find that, as far as local marriages are concerned, in all the analyzed nations the Brock and Durlauf scenario (discrete choice wit social interactions) prevails: at the national level, the relation between local marriages and potential couples always follows a square root, apart for the case of Italy. In the latter, two well-separated square root growths appear and mirror the marriage evolution of two detached communities that, remarkably, do coincide with the Northern and Southern regional clusters into which the peninsula were split before its unification in $1861$. This finding may suggest that Italy still experiences persistence of lasting heritages of different cultures that have not mixed yet.
\newline
Inspired by the {\em Italian case}, to deepen the possible existence of regional clusters (i.e. ensembles of {\em geographically adjacent} regions sharing close behaviors)  within each analyzed country, we then moved to study data at the regional level, collecting the trend in their marriage evolution for each region and then comparing regional behavior with the averaged-national one. In this way it was possible to distinguish  regions where the evolution of marriages is, respectively, consistent with, overestimated, or underestimated by the average (i.e., at the country level) behavior. Interestingly, we also find that regions with analogous outcomes typically form connected clusters. While this effect is mild for France and Spain, it becomes manifest for Italy, highlighting the net presence of behavioral differences between its Northern and Southern regions.
Such heterogeneity is absent in Germany and Switzerland, where the behavior of all the regions, within the experimental error, fall into the main class paved by the national reference.

Moving to mixed marriages, we found that the Brock and Durlauf scenario becomes much less pronounced, possibly surviving only in Spain (for low values of possible mixed couples) and in Switzerland (which is somehow a case by itself given that its immigration policies do not have to overlap with EU prescriptions).
\newline
For all the other countries we found that it is the Mc Fadden theory to reproduce remarkably well the behavior (i.e. in France and Germany the relation between marriages and potential -mixed- couples follows a sharp straight line, while in Italy noise in the data forbids to make sharp statements).  This possibly suggests that imitational mechanisms, widespread among decision makers within modern societies, may require a higher level of integration when migrants are concerned as their establishment in these mixed cases seems not to have taken place yet.

\begin{acknowledgments}
The authors are grateful to an anonymous referee for pointing out interesting historical similarities in the socio-economical growth of Germany and Italy during the past century.
\newline
INdAM-GNFM (Progetto Giovani Agliari2014), EPSRC  support and Sapienza Universit\`a di Roma are acknowledged.
\newline
AB is grateful to LIFE group (Laboratories for Information, Food and Energy) for partial financial support through  {\em programma INNOVA, progetto MATCH}.
\newline
MAJ would like to thank Fondazione Banco di Sardegna for supporting his work.
\newline
AP acknowledges support by the Engineering and Physical Sciences Research Council (EPSRC), Grant No. EP/L505110/1.
\newline
Competing financial interests: the authors declare no competing financial interests.

\end{acknowledgments}

%

\end{document}